\documentclass[%
 reprint,
 amsmath,amssymb,
 aps,
floatfix,
longbibliography
]{revtex4-2}

\usepackage{graphicx}
\usepackage{changepage}
\usepackage{dcolumn}
\usepackage{chemformula}
\usepackage{gensymb}
\usepackage{xfrac}
\usepackage{parskip}
\usepackage{makecell}
\usepackage{subfigure}
\usepackage{float}
\usepackage{bm}
\usepackage{natbib}
\usepackage{threeparttablex}

\renewcommand{\arraystretch}{1.5}

\begin{document}

\setlength{\parindent}{20pt} 
\setlength{\parskip}{0pt}

\preprint{APS/123-QED}

\title{Magnetic properties of \ch{RE_{2}O_{2}CO_{3}} (RE = Pr, Nd, Gd, Tb, Dy, Ho, Er, Yb) with a rare earth-bilayer of triangular lattice}
        
\author{Aya Rutherford}
   \affiliation{Department of Physics and Astronomy, University of Tennessee, Knoxville, TN 37996, USA}
\author{Chengkun Xing}
   \affiliation{Department of Physics and Astronomy, University of Tennessee, Knoxville, TN 37996, USA}   
\author{Qing Huang}
   \affiliation{Department of Physics and Astronomy, Louisiana State University, Baton Rouge, Louisiana 70803, USA} 
\author{Eun Sang Choi}
   \affiliation{National High Magnetic Field Laboratory, Florida State University, Tallahassee, FL 32310, USA}
\author{Stuart Calder}
   \affiliation{Neutron Scattering Division, Oak Ridge National Laboratory, Oak Ridge, Tennessee 37831, USA}
\author{Haidong Zhou}
   \email{hzhou10@utk.edu}
   \affiliation{Department of Physics and Astronomy, University of Tennessee, Knoxville, TN 37996, USA}
 
\date{\today}

\begin{abstract}
     Polycrystalline samples of \ch{RE_{2}O_{2}CO_{3}} (RE = Pr, Nd, Gd, Tb, Dy, Ho, Er, and Yb) with a unique rare-earth bilayer of triangular lattice were synthesized and studied by DC and AC magnetic susceptibility. Data reveals various magnetic ground states including (i) a nonmagnetic ground state for the Pr sample; (ii) long range magnetic ordering for the Nd, Gd, Tb, Dy, Ho, and Er samples. Besides the Gd sample, they exhibit field-induced spin state transitions. More interestingly, the series spin state transitions in the Nd and Dy samples could be attributed to the field-induced up-up-down (UUD) spin structure. Neutron powder diffraction (NPD) measurements of the Er sample suggest a spiral spin structure below its $T_{\text{N}}$; and (iii) a short range ordering for the Yb sample.  The disrupted inter-layer interaction due to the shift of \ch{Yb^{3+}} ions within the bilayer prevents long range magnetic ordering down to 30 mK and makes it another Yb-related triangular lattice antiferromagnet that has the potential to realize a quantum spin liquid state. 
     
    \noindent{DOI:}
\end{abstract}

\maketitle

\begin{figure*}[htb!]
    \centering
    \begin{minipage}[b]{0.5\linewidth}
        \centering
        \includegraphics[width=1.0\linewidth]{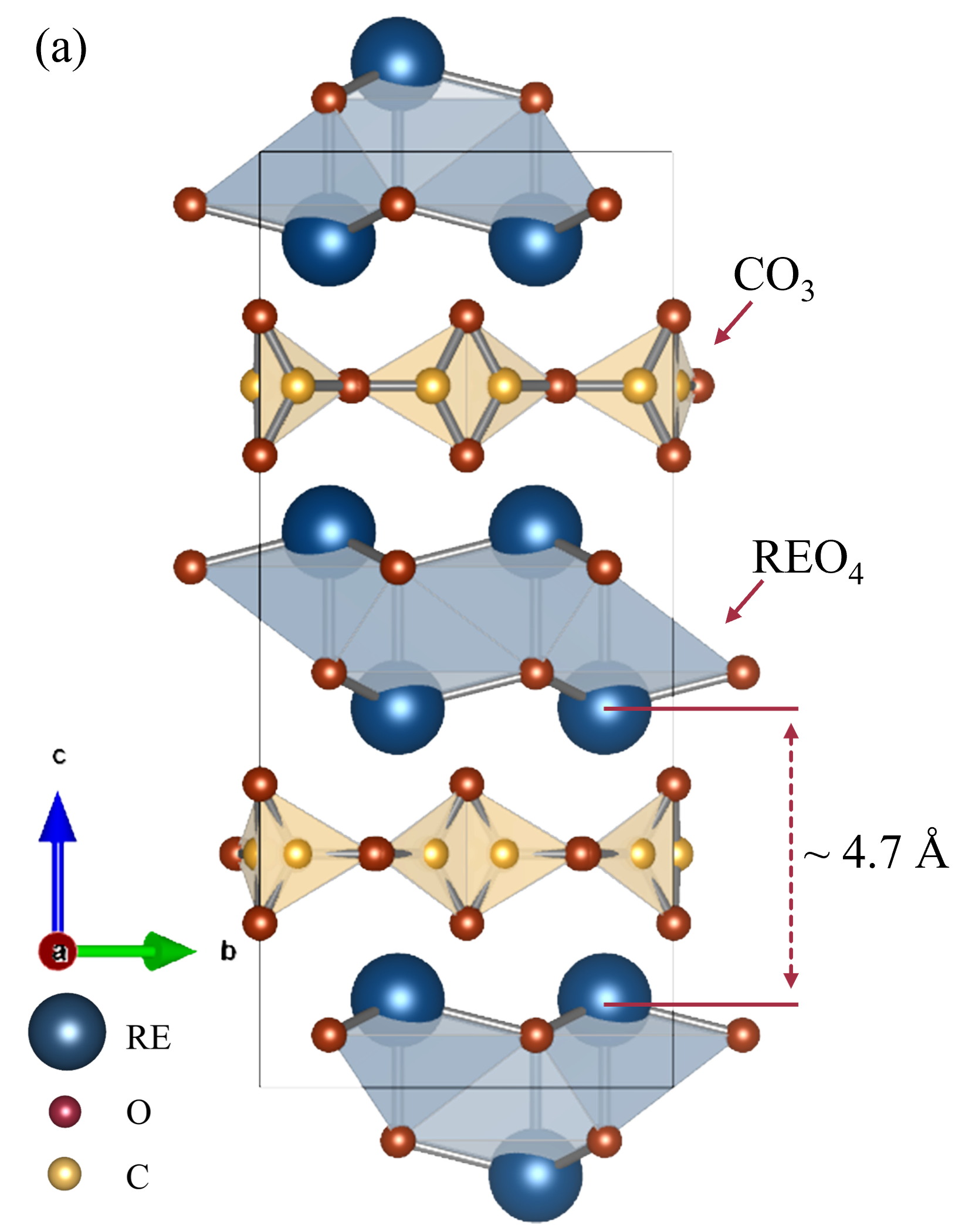}
    \end{minipage}%
    \begin{minipage}[b]{0.4\linewidth}
        \includegraphics[width=0.9\linewidth]{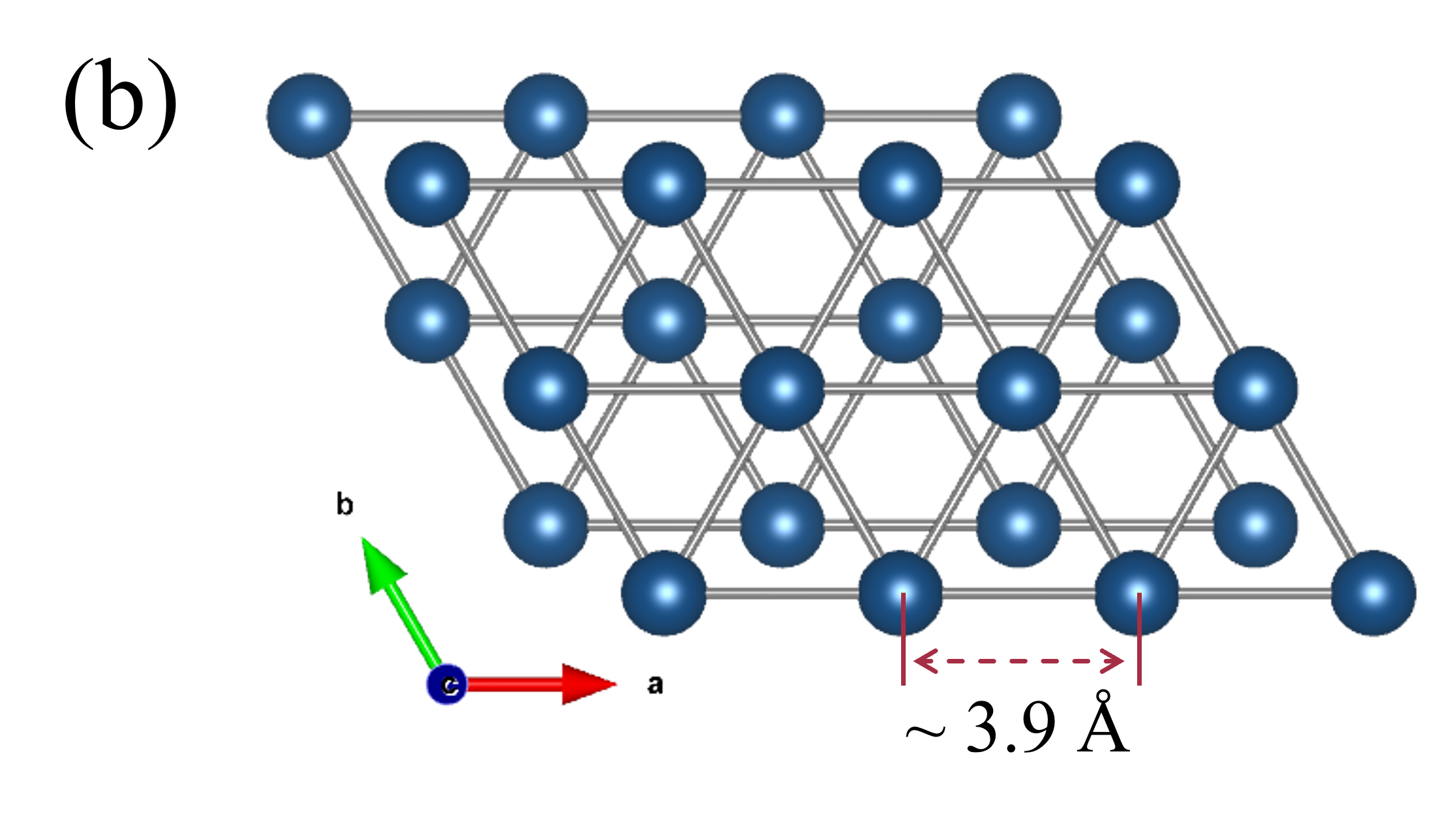}
        \hbox to \linewidth{} 
        \includegraphics[width=0.85\linewidth]{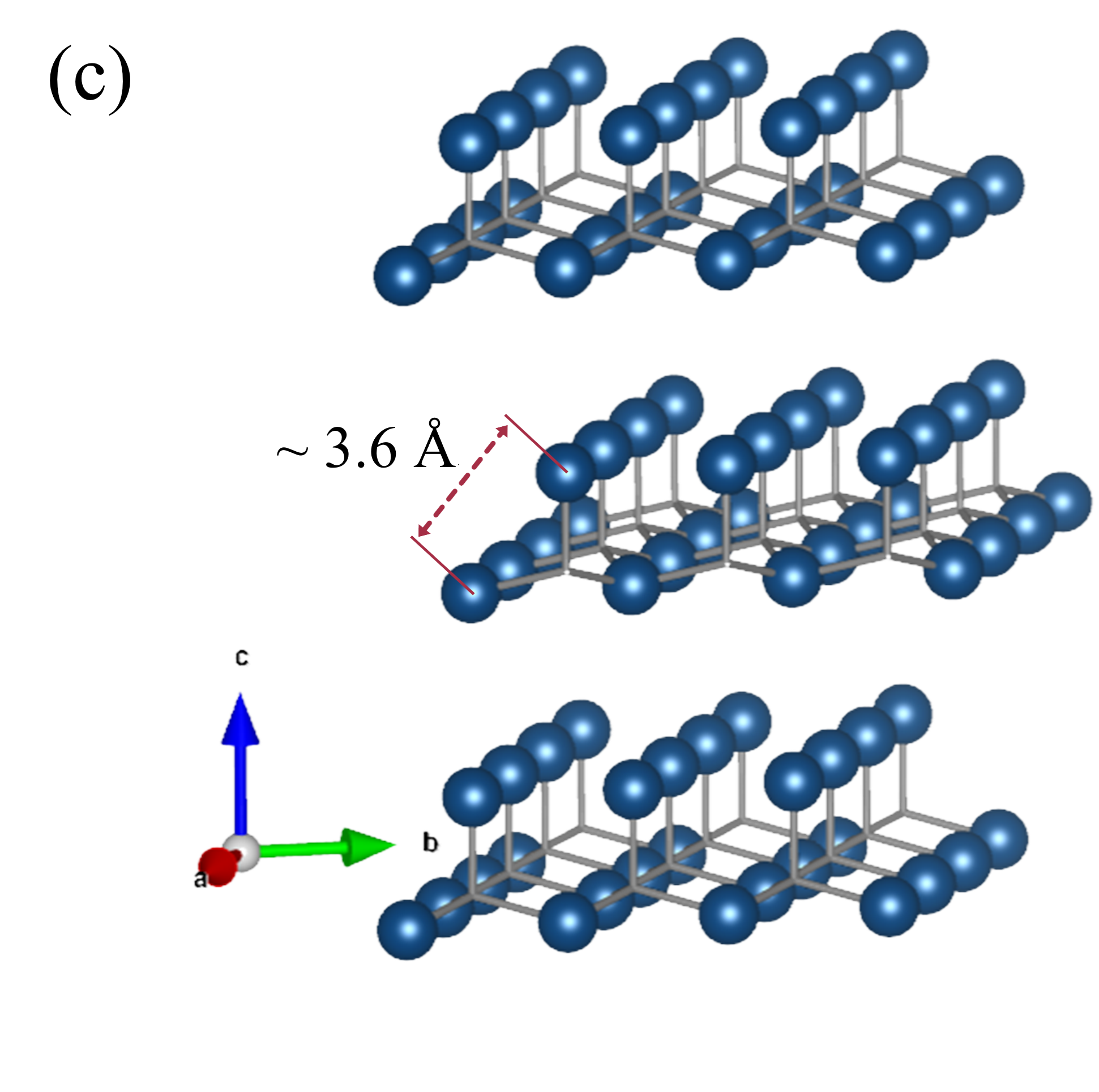}
    \end{minipage}
    \caption{(a) Crystal structure of \ch{RE_{2}O_{2}CO_{3}}. (b) Arrangement of RE$^{3+}$ ions in a triangular lattice bilayer. (c) Arrangement of RE$^{3+}$ ions in shifted triangular lattice layers along the $c$-axis.}
    \label{fig:struct}
\end{figure*}

\begin{figure*}[p]
    \includegraphics[width=2.0\columnwidth]{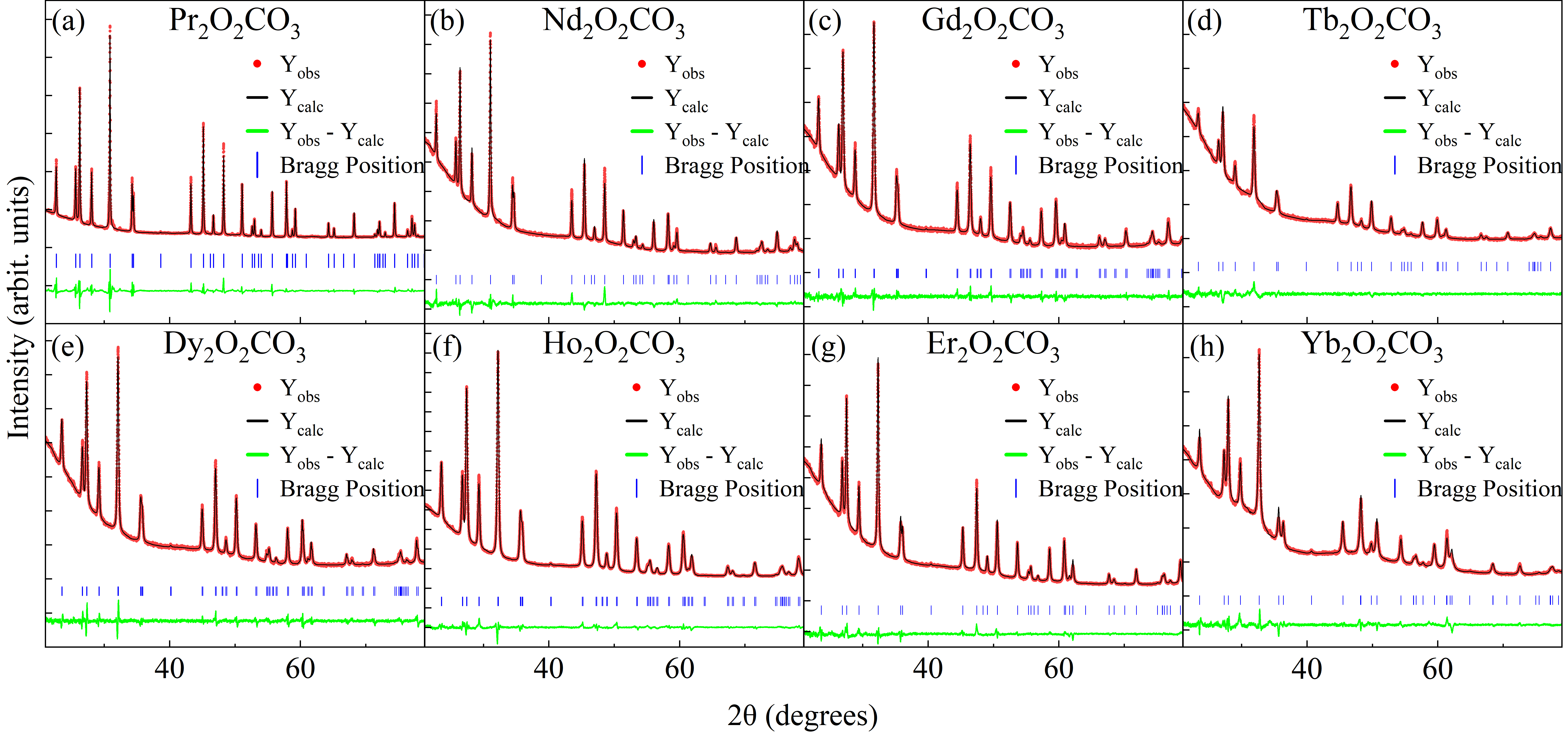}
    \caption{Room temperature powder XRD patterns and best fits from Rietveld refinement.}
    \label{fig:rr}
\end{figure*}

\begin{table*}[p]
    \renewcommand{\arraystretch}{1.5}
    \caption{\label{tab:table1}Summary of room-temperature XRD Rietveld refinements for \ch{RE_{2}O_{2}CO_{3}}.}
    \label{table:rr}
    \begin{ruledtabular}
    \begin{tabular}{ccccccccc}
        RE & Pr & Nd& Gd & Tb & Dy  & Ho& Er& Yb \\
        \hline
        \ch{RE^{3+}} IR (\AA)& 1.126& 1.109& 1.053& 1.040& 1.027& 1.015& 1.004&  0.985\\
        \multicolumn{1}{r}{$a$} (\AA) & 4.01240(1)& 3.98902(1)& 3.90984(1)& 3.88567(2)&  3.86389(1)& 3.84738(1)&  3.82963(1)&  3.76977(2)\\
        \multicolumn{1}{r}{$c$} (\AA) & 15.69711(4)& 15.61398(8)& 15.31344(9)& 15.23047(12)&  15.14404(10)& 15.11306(6)&  15.03997(6)&  15.09768(12)\\
        RE & (\(\frac{1}{3}\), \(\frac{2}{3}\), z) & (\(\frac{1}{3}\), \(\frac{2}{3}\), z) & (\(\frac{1}{3}\), \(\frac{2}{3}\), z) & (\(\frac{1}{3}\), \(\frac{2}{3}\), z) &  (\(\frac{1}{3}\), \(\frac{2}{3}\), z) & (\(\frac{1}{3}\), \(\frac{2}{3}\), z) &  (\(\frac{1}{3}\), \(\frac{2}{3}\), z) &  (\(\frac{1}{3}\), \(\frac{2}{3}\), z) \\
        \multicolumn{1}{r}{z}  & 0.09484(3)& 0.09496(3)& 0.09417(3)& 0.09415(4)&  0.09397(4)& 0.09358(2)&  0.09352(2)&  0.09242(4)\\
        C & (x, 2x, \(\frac{1}{4}\))& (x, 2x, \(\frac{1}{4}\))& (x, 2x, \(\frac{1}{4}\))& (x, 2x, \(\frac{1}{4}\))&  (x, 2x, \(\frac{1}{4}\))& (x, 2x, \(\frac{1}{4}\))&  (x, 2x, \(\frac{1}{4}\))&  (x, 2x, \(\frac{1}{4}\))\\
        \multicolumn{1}{r}{x}  & 0.06006(8)& 0.42056(4)& 0.06017(3)& 0.07208(1)& 0.06850(6)& 0.02017(6)&  0.08791(5)&  0.08841(5)\\
        \multicolumn{1}{r}{y}  & 0.12021(16)& 0.84114(8)& 0.12048(6)& 0.14429(2)&  0.13710(12)& 0.04041(12)&  0.17591(10)&  0.17692(10)\\
        $\text{O}_{1}$& (\(\frac{1}{3}\), \(\frac{2}{3}\), z) & (\(\frac{1}{3}\), \(\frac{2}{3}\), z) & (\(\frac{1}{3}\), \(\frac{2}{3}\), z) & (\(\frac{1}{3}\), \(\frac{2}{3}\), z) &  (\(\frac{1}{3}\), \(\frac{2}{3}\), z) & (\(\frac{1}{3}\), \(\frac{2}{3}\), z) &  (\(\frac{1}{3}\), \(\frac{2}{3}\), z) &  (\(\frac{1}{3}\), \(\frac{2}{3}\), z) \\
        \multicolumn{1}{r}{z}  & 0.55722(21)& 0.55477(21)& 0.55661(19)& 0.55351(24)&  0.55628(22)& 0.55698(10)&  0.55638(17)& 0.55285(27)\\
        $\text{O}_{2}$ & (0, 0, z) & (0, 0, z) & (0, 0, z) &(0, 0, z) &  (0, 0, z) & (0, 0, z) &  (0, 0, z) &  (0, 0, z) \\
        \multicolumn{1}{r}{z}  & 0.17920(35)& 0.17358(36)& 0.16921(38)& 0.17357(51)&  0.18052(34)& 0.17474(17)&  0.18251(26)&  0.17653(37)\\
        $\text{O}_{3}$ & (x, 2x, \(\frac{1}{4}\))& (x, 2x, \(\frac{1}{4}\))& (x, 2x, \(\frac{1}{4}\))& (x, 2x, \(\frac{1}{4}\))&  (x, 2x, \(\frac{1}{4}\))& (x, 2x, \(\frac{1}{4}\))&  (x, 2x, \(\frac{1}{4}\))&  (x, 2x, \(\frac{1}{4}\))\\
        \multicolumn{1}{r}{x}  & 0.20692(2)& 0.10443(2)& 0.23751(1)& 0.23751(1)&  0.23909(3)& 0.23686(2)&  0.24315(2)&  0.21784(5)\\
        \multicolumn{1}{r}{y}  & 0.41385(5)& 0.20886(4)& 0.46943(3)& 0.47502(3)&  0.47818(6)& 0.47371(4)&  0.48629(4)&  0.43568(10)\\
        $\text{R}_\text{p}$ & 2.77& 1.26& 1.05& 1.16&  1.19& 1.59&  1.36 &  1.49\\
        $\text{R}_{\text{wp}}$ & 4.20& 1.90& 1.45& 1.59&  1.77& 2.28&  1.96 &  2.14\\
        $\chi^2$ & 20.3& 4.20& 3.28& 3.06&  4.39& 5.64&  3.94 &  4.92\\
    \end{tabular}
    \end{ruledtabular}
\end{table*}

\begin{figure}[htb!]
    \centering
    \includegraphics[width=1.0\columnwidth]{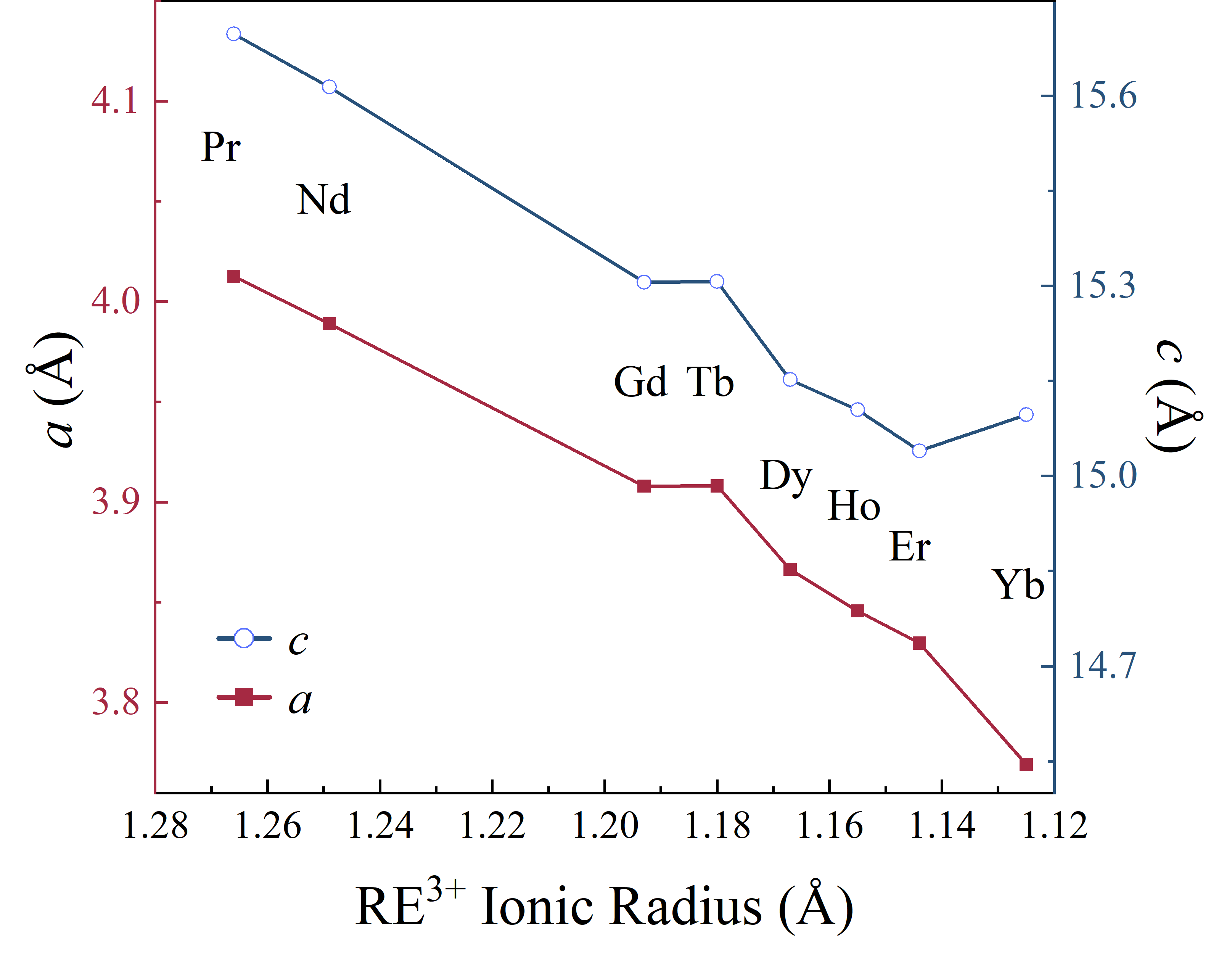}
    \caption{\label{fig:epsart} Lattice parameters obtained from Rietveld refinements as a function of the \ch{RE^{3+}} ionic radius.}
    \label{fig:param}
\end{figure}

\section{\label{sec:level1}Introduction}

    Rare-earth triangular lattice antiferromagnets (RE-TLAF) have attracted great interest in recent years. This fervor can be attributed to two of their characteristic elements: first, the triangular lattice, which is one of the simplest geometrically-frustrated lattices \cite{anderson:1973,ramirez:1996,collins:1997,harrison:2004,balents:2010}. Second, the large spin-orbit coupling (SOC) and crystal electric field (CEF) effects of the RE ions' 4$f$ electrons, which can lead to highly anisotropic exchange interactions \cite{moriya:1960,arjun:2016}. This combination makes RE-TLAFs an excellent platform for exploring exotic magnetic properties.
    
    One driving force in the study of RE-TLAFs is the search for the quantum spin liquid (QSL), a highly spin-entangled state that is unable to attain magnetic ordering down to ultra-low temperatures \cite{balents:2010,zhou:2017,savary:2017,li:2020IP,broholm:2020,chamorro:2020}. Most efforts have been focused on Ytterbium-based TLAFs as their easy-plane anisotropy and the $J_{\text{eff}}$ = \sfrac{1}{2} ground state of \ch{Yb^{3+}} ions have the potential to realize the QSL state. Indeed, several Yb-TLAFs, including \ch{NaYbO_{2}} \cite{hashimoto:2003,ranjith:2019PRB,ding:2019,bordelon:2019,bordelon:2020,jguo:2020}, \ch{NaYbS_{2}} \cite{baenitz:2018,jia:2023,sarkar:2019}, \ch{NaYbSe_{2}} \cite{ranjith:2019,lin:2023,zhang:2021,zhang:2022,dai:2021,xu:2022,jia:2020}, \ch{CsYbSe_{2}} \cite{xing:2020QM,sereika:2023,xing:2019,pai:2022,xie:2023}, \ch{YbBO_{3}} \cite{somesh:2023,sala:2023,mukherjee:2018}, \ch{YbMgGaO_{4}} \cite{rao:2021,li:2015PRL,li:2015Nat,paddison:2016,li:2017,shen:2016}, \ch{YbZn_{2}GaO_{5}} \cite{xu:2023}, \ch{K_{3}Yb(VO_{4})_{2}} \cite{voma:2021}, and \ch{Ba_{6}Yb_{2}Ti_{4}O_{17}} \cite{song:2023,khatua:2024}, have been studied as QSL candidates. More recently, it was noticed that dominant Ising-like exchange interactions also can lead to a QSL state in RE-TLAFs, such as in \ch{NdTa_{7}O_{19}} \cite{arh:2022}, \ch{CsNdSe_{2}} \cite{xing:2023,xing:2020}, \ch{KTmSe_{2}} \cite{zheng:2023}, and \ch{PrBAl_{11}O_{19}} (B = Mg, Zn) \cite{bu:2022,zhen:2024}. The \ch{Tm^{3+}} and \ch{Pr^{3+}} ions of \ch{KTmSe_{2}} and \ch{PrBAl_{11}O_{19}} are non-Kramers ions whose doublet ground states usually split into two singlets. The spin-spin interaction can serve as a local exchange field on top of the double singlets. Therefore, their magnetic properties can be mapped to an Ising model in a transverse magnetic field. Moreover, while the magnetic exchange interaction is weak, the dipolar interaction of RE ions could also drive a system to a liquid-like state in RE-TLAFs, such as in \ch{NaBaYb(BO_{3})_{2}} \cite{guo:2019}, \ch{KBaYb(BO_{3})_{2}} \cite{sanders:2017,pan:2021}, and \ch{Ba_{3}Yb(BO_{3})_{3}} \cite{bag:2021}.

    Other interesting topics associated with RE-TLAFs include (i) proximate QSL behavior in a system with low temperature long-range magnetic ordering, for example in \ch{KYbSe_{2}} \cite{scheie:2024,scheie:2024PRB}; (ii) the spin liquid state and topological structural defects in hexagonal \ch{TbInO_{3}} \cite{kim:2019,jkim:2019,clark:2019,ye:2021,jung:2023}; (iii) the Berezinskii-Kosterlitz-Thouless (BKT) phase with a quantum critical point, which has been proposed for \ch{TmMgGaO_{4}} \cite{cevallos:2018,shen:2019,liu:2020,hu:2020,li:2020,dun:2021,frandsen:2022} and \ch{Ba_{3}Gd(BO_{3})_{3}} \cite{kelly:2020}; (iv) long range magnetic ordered states with rare spin structures, such as the stripe order observed in \ch{CsCeSe_{2}} \cite{xing:2020} and \ch{KErSe_{2}} \cite{xing:2021PRB} and the realization of a two-sublattice exchange in \ch{Ba_{3}Er(BO_{3})_{3}} \cite{ennis:2024}; and, (v) the interplay between the frustrated magnetism and charge carriers in metallic \ch{CeCd_{3}P_{3}} \cite{avers:2021,lee:2019} and semiconductor \ch{EuCd_{2}As_{2}} \cite{jo:2020,xu:2021,wang:2021,sun:2022,santos-cottin:2023}. 

    \ch{RE_{2}O_{2}CO_{3}} \cite{turcotte:1968,kutlu:1999,christensen:1970,olafsen:2001} is a family of compounds with rare earth triangular lattice bilayers that has been little studied as of yet. Magnetic properties of RE = Nd, Sm, and Dy have been briefly reported but their magnetic phase diagrams have not been described \cite{arjun:2016}. Meanwhile, the magnetic properties of the other RE members are still unknown. In this paper, we synthesized and reported the magnetic ground states and magnetic phase diagrams of eight \ch{RE_{2}O_{2}CO_{3}} members (RE = Pr, Nd, Gd, Tb, Dy, Ho, Er, and Yb) by measuring DC and AC magnetic susceptibility. We also performed neutron powder diffraction (NPD) measurements on the Er sample.
    
\begin{figure*}[htb!]
    \centering
    \includegraphics[width=2.0\columnwidth]{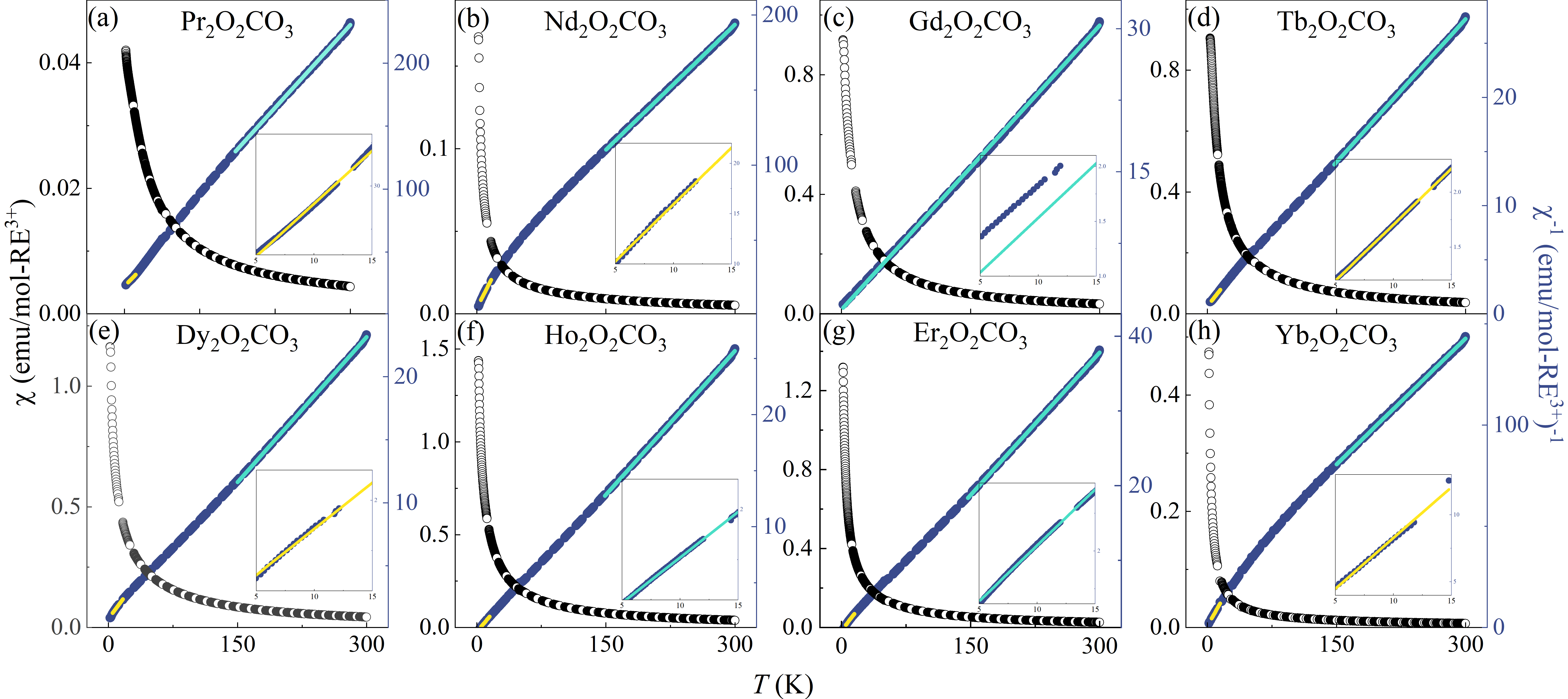}
    \caption{DC susceptibility $\chi$ (left axis) and its inverse $\chi^{-1}$ (right axis) from 2 to 300 K at 0.1 T for \ch{RE_{2}O_{2}CO_{3}}. High temperature (150--300 K) fits of inverse susceptibility are blue, low temperature (5--15 K) fits are yellow. Insets: $\chi^{-1}$ at low-temperature regions (5--15 K). The Curie-Weiss fit of \ch{Gd_{2}O_{2}CO_{3}} was instead over the entire temperature regime (2--300 K) in blue, with inset of the low temperature region.}
    \label{fig:sus}
\end{figure*}

\begin{figure*}[htb!]
    \centering
    \includegraphics[width=2.0\columnwidth]{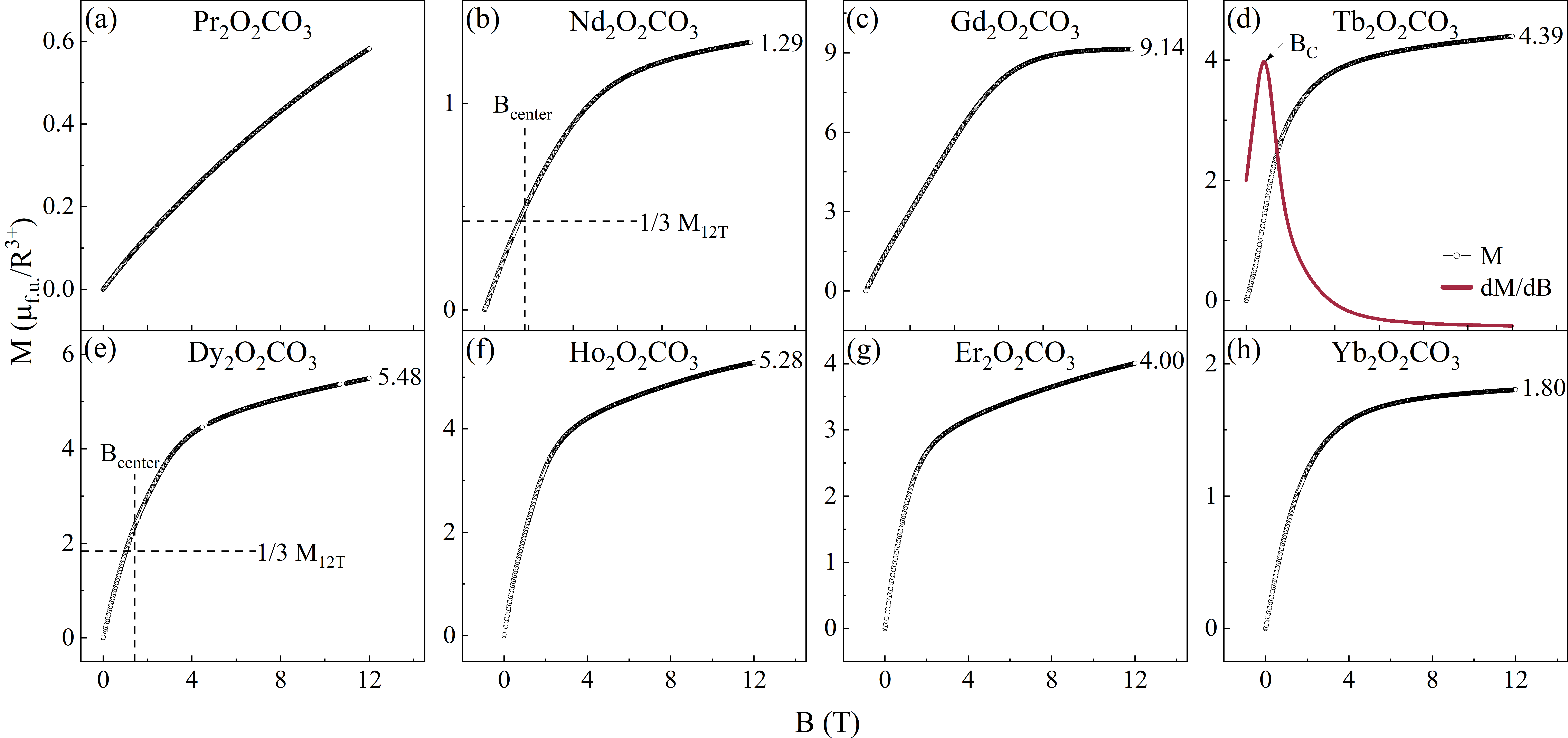}
    \caption{DC magnetization curves measured at 2 K from 0 to 12 T for \ch{RE_{2}O_{2}CO_{3}}. The estimated saturation values, $\text{M}_{\text{12T}}$, are labeled at the end of each curve besides the Pr sample. The derivative of \ch{Tb_{2}O_{2}CO_{3}} is plotted in red. The curves of \ch{Nd_{2}O_{2}CO_{3}} and \ch{Dy_{2}O_{2}CO_{3}} include demarcations of one-third of the $\text{M}_{\text{12T}}$ value and $\text{B}_{\text{center}}$ from $\chi'$(B) data.}
    \label{fig:mag}
\end{figure*}

\begin{table*}[htb!]
    \caption{\label{table:magprop}Summary of magnetism parameters for \ch{RE_{2}O_{2}CO_{3}}.}
    \renewcommand{\arraystretch}{1.5}
    \begin{ruledtabular}
    \begin{tabular}{cccccccc}
        Compound & \makecell{Low \textit{T} \\ $\theta_{\text{CW}}$ (K)}& \makecell{Low \textit{T} \\ $\mu_{\text{eff}}^{\text{exp}}$ ($\mu_{\text{B}}$)} & \makecell{High \textit{T} \\ $\theta_{\text{CW}}$ (K)} & \makecell{High \textit{T} \\ $\mu_{\text{eff}}^{\text{exp}}$ ($\mu_{\text{B}}$)}& \makecell{$\mu_{\text{eff}}^{\text{theo}}$ ($\mu_{\text{B}})$}  & \makecell{$\text{M}_{\text{12T}}$ \\ (\ch{$\mu_{\text{B}}$/RE^{3+}}) } &\makecell{$\text{M}_{\text{S}}$  (\ch{$\mu_{\text{B}}$/RE^{3+}}) }\\
        \hline
        \ch{Pr_{2}O_{2}CO_{3}}  & -37.32& 3.62& -49.71& 3.48& 3.58  &-- &3.20\\
        \ch{Nd_{2}O_{2}CO_{3}}  & -4.13& 2.66& -48.51& 3.79& 3.62  &1.295 &3.27\\
        \ch{Gd_{2}O_{2}CO_{3}}  & ---& ---& -5.42& -8.97& 7.94  &9.14 &7.00\\
        \ch{Tb_{2}O_{2}CO_{3}}  & -6.70& 8.85& -9.06& 9.52& 9.72 &4.39 &9.00\\
        \ch{Dy_{2}O_{2}CO_{3}}  & -8.70& 9.33&  -1.30& 10.20& 10.65  &5.48 &10.0\\
        \ch{Ho_{2}O_{2}CO_{3}}  & -6.83& 9.41& -0.19& 9.67& 10.61  &5.28 &10.0\\
        \ch{Er_{2}O_{2}CO_{3}}  & -4.00& 7.35& -2.94& 7.85& 9.58  &4.00 &9.00\\
        \ch{Yb_{2}O_{2}CO_{3}}  & -0.94& 3.25& -44.25& 4.39& 4.54  &1.80 &4.00\\
    \end{tabular}
    \end{ruledtabular}
\end{table*}
    
\section{\label{sec:level2}Experimental Details}

    Polycrytalline samples of \ch{RE_{2}O_{2}CO_{3}} (RE = Pr, Nd, Gd, Tb, Dy, Ho, Er, Yb) were synthesized in a two-step process. First, appropriate amounts of \ch{RE(NO_{3})_{3} $\cdot X$ H_{2}O} and \ch{NaHCO_{3}} were dissolved in water to form a white precipitate. Then, this precipitate was filtered and annealed in a vacuumed and sealed quartz tube  at temperatures of 550\degree C, 500\degree C, 375\degree C, 305\degree C, 370\degree C, 360\degree C, 340\degree C, 295\degree C for Pr, Nd, Gd, Tb, Dy, Ho, Er, Yb, respectively, for 24 to 36 hours.
    
    The phase purity of the resulting polycrystalline samples was checked via room-temperature powder x-ray diffraction (XRD) using a HUBER imaging plate Guinier camera 670 with Cu radiation ($\lambda$ = 1.54059 \AA{}). The XRD refinements were performed with the FULLPROF program. The crystal structures were plotted and the bond distances were computed using VESTA.  DC magnetic susceptibility and magnetization were measured in a Physical Property Measurement System (PPMS) from Quantum Design with the vibrating sample magnetometer (VSM) option equipped. The DC magnetic susceptibility was measured at a magnetic field B = 0.1 T from 2 to 300 K and the DC magnetization was measured at 2 K from 0 to 12 T. The AC susceptibility measurements were conducted with a voltage-controlled current source (Stanford Research, Model CS580) and lock-in amplifier (Stanford Research, Model SR830) \cite{dun:2014}. The RMS amplitude of the AC excitation field was set to be 0.6 Oe with the frequency between 80 Hz to 2147 Hz. The measurements were performed at SCM1 and SCM2 of the National High Magnetic Field Laboratory, Tallahassee. 

    The neutron powder diffraction (NPD) measurement was performed at the Neutron Powder Diffractometer (HB-2A) at HFIR, ORNL to study the nuclear and magnetic structure of \ch{Er_{2}O_{2}CO_{3}} \cite{calder:2018}. A powder sample weighing approximately 0.5 grams was prepared for the experiment. The nuclear structure was studied by NPD measurements with 1.54 \AA{} neutron wavelength to ensure a larger Q coverage. Further measurements with a 2.41 \AA{} neutron wavelength were utilized to focus on the magnetic signal in the low Q range. The diffraction data was analyzed by the Rietveld refinement technique within the FULLPROF program package.
\section{\label{sec:level3}Structure}

    All members of the \ch{RE_{2}O_{2}CO_{3}} family are isostructural and crystallize into the hexagonal space group $\text{P}6_{3}/\text{mmc}$. There are several distinct structural features. First, the corner-sharing \ch{REO_{4}} tetrahedra network in the $ab$ plane forms an isotropic RE-triangular layer (Fig. \ref{fig:struct}). Second, each \ch{REO_{4}} tetrahedron is edge-sharing with three other tetrahedra of the adjacent layer along the $c$-axis, thus forming a RE-triangular lattice bilayer. The bilayers are separated from each other by carbonate groups along the $c$-axis. Third, very importantly, within each bilayer, the two RE-triangular lattices are displaced with respect to each other in such a way that the RE ion in one layer is projected towards the center of the triangle formed by the RE ions in the adjacent layer along the $c$ axis, as shown in Fig. \ref{fig:struct}(b, c).

    The XRD patterns for \ch{RE_{2}O_{2}CO_{3}} (RE = PR, Nd, Gd, Tb, Dy, Ho, Er, and Yb) with the refinement results superimposed on the data are shown in Fig. \ref{fig:rr} and the experimental patterns agree well with the calculated ones. The structural refinement results are provided in Table \ref{table:rr}. Figure \ref{fig:param} shows the lattice parameters $a$ (left axis) and $c$ (right axis) versus ionic radius (IR) of the \ch{RE^{3+}} ions. The lattice constants decrease approximately linearly with decreasing ionic radius. One anomaly is \ch{Yb_{2}O_{2}CO_{3}}, which shows an unexpected larger $c$ lattice parameter. Generally, the nearest neighbor distance for RE ions on a triangular layer is around 3.9 \AA{}, the nearest distance between RE ions on adjacent RE-triangular layers within each bilayer is around 3.6 \AA{}, and the distance between two adjacent bilayers along the $c$-axis is around 4.7 \AA{}, as plotted in Fig. \ref{fig:struct}.

\section{\label{sec:level4}Magnetic Properties}

    The DC magnetic susceptibility and DC magnetization data of all eight samples are shown in Fig. \ref{fig:sus} and \ref{fig:mag}, respectively. The Curie Weiss (CW) law, \sfrac{1}{$\chi$} = $\frac{\textit{T} - \theta_{CW}}{C}$ was used to fit the inverse susceptibility at both the high temperature regime [150 - 300 K] and low temperature regime [5 - 15 K]. The obtained Curie-Weiss temperatures $\theta_{\text{CW}}$, and effective moments $\mu_{\text{eff}}$ are listed in Table \ref{table:magprop} with the theoretically calculated free ion moment for comparison. The theoretical saturation moment \ch{M_{s}} = $g_{\text{theo}}$$J$, where $g_{\text{theo}}$ is the expected $g$ factor for free-ions and  $J$ is the theoretical Hund's value, and the experimental saturation value at 12 T, \ch{M_{12T}}, are also listed in Table \ref{table:magprop}. Here, we assume that the magnetization value at 12 T (\ch{M_{12T}}) is approaching saturation. Due to the linear increase of the magnetization at high field (Fig. \ref{fig:mag}), it is difficult to exactly extract the saturation moment. The description of the DC magnetic properties and AC susceptibility data for each sample is listed below.

    \subsection{\ch{Pr_{2}O_{2}CO_{3}}}
    
        For \ch{Pr_{2}O_{2}CO_{3}}, the high temperature CW fit of \sfrac{1}{$\chi$} (Fig. \ref{fig:sus}(a)) yields a $\theta_{\text{CW}}$ = -49.71 K and a $\mu_{\text{eff}}$ = 3.48 $\mu_{\text{B}}$. This $\mu_{\text{eff}}$ value is consistent with the free-ion moment of $\mu_{\text{eff}}^{\text{theo}}$ = 3.58 $\mu_{\text{B}}$ expected for \ch{Pr^{3+}} ions. \sfrac{1}{$\chi$} shows a weak slope change below 50 K, suggesting changes of magnetic moments and spin-spin interactions in these temperature regions due to CEF effects. The low temperature CW fit yields a $\theta_{\text{CW}}$ = -37.32 K and a $\mu_{\text{eff}}$ = 3.62 $\mu_{\text{B}}$. Such large $\theta_{\text{CW}}$ does not necessarily reflect the intrinsic strength of the magnetic exchange interaction of \ch{Pr_{2}O_{2}CO_{3}}, but could be related to the competition between the exchange interaction and the split between the ground-state doublet of \ch{Pr^{3+}} ions (see details in the Discussion). In fact, our AC susceptibility measurement shows weak and noisy signals (not plotted here), which suggests a nonmagnetic ground state. 
                
    \begin{figure*}[htb!]
        \centering
        \includegraphics[width=0.67\columnwidth]{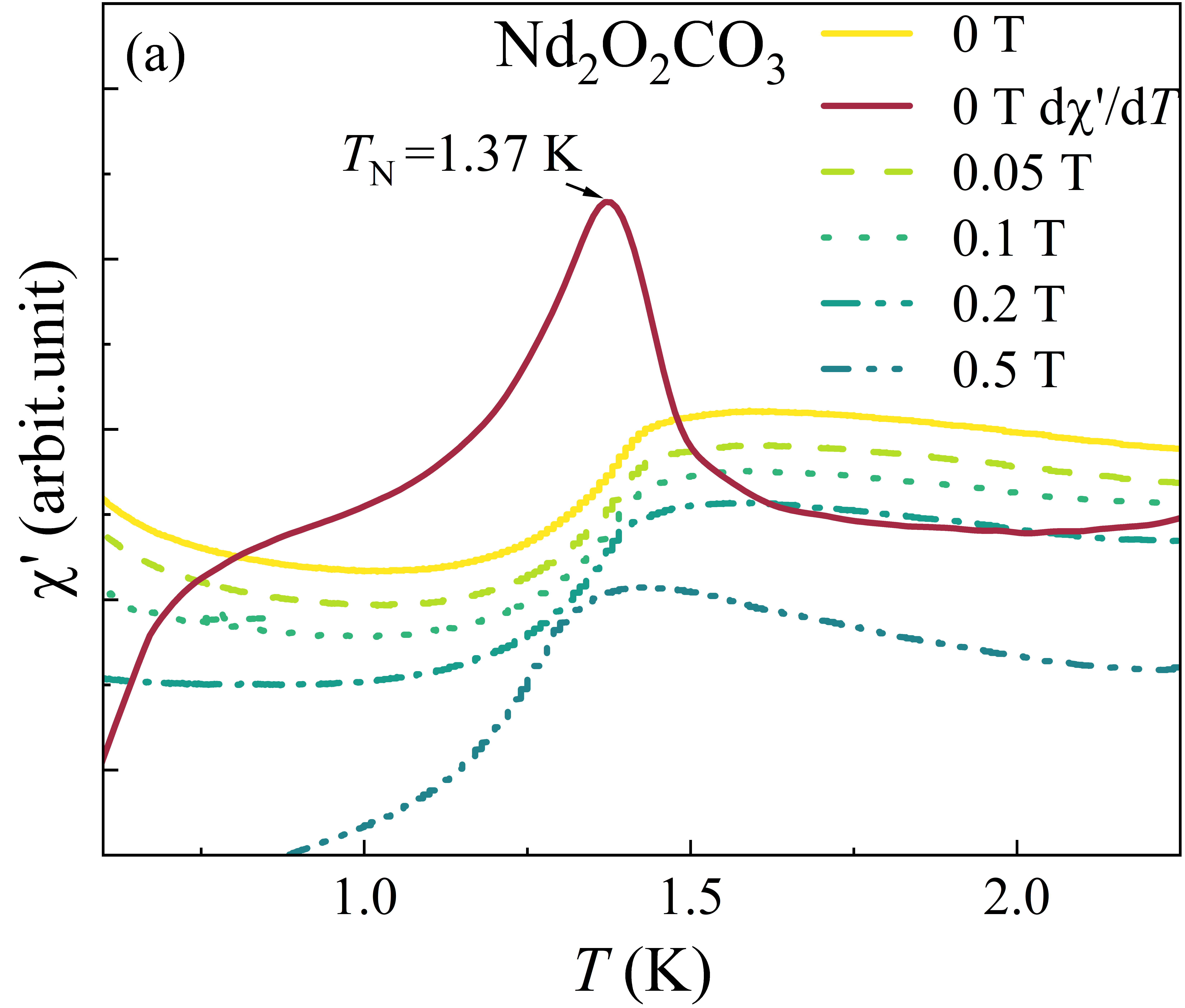}
        \includegraphics[width=0.67\columnwidth]{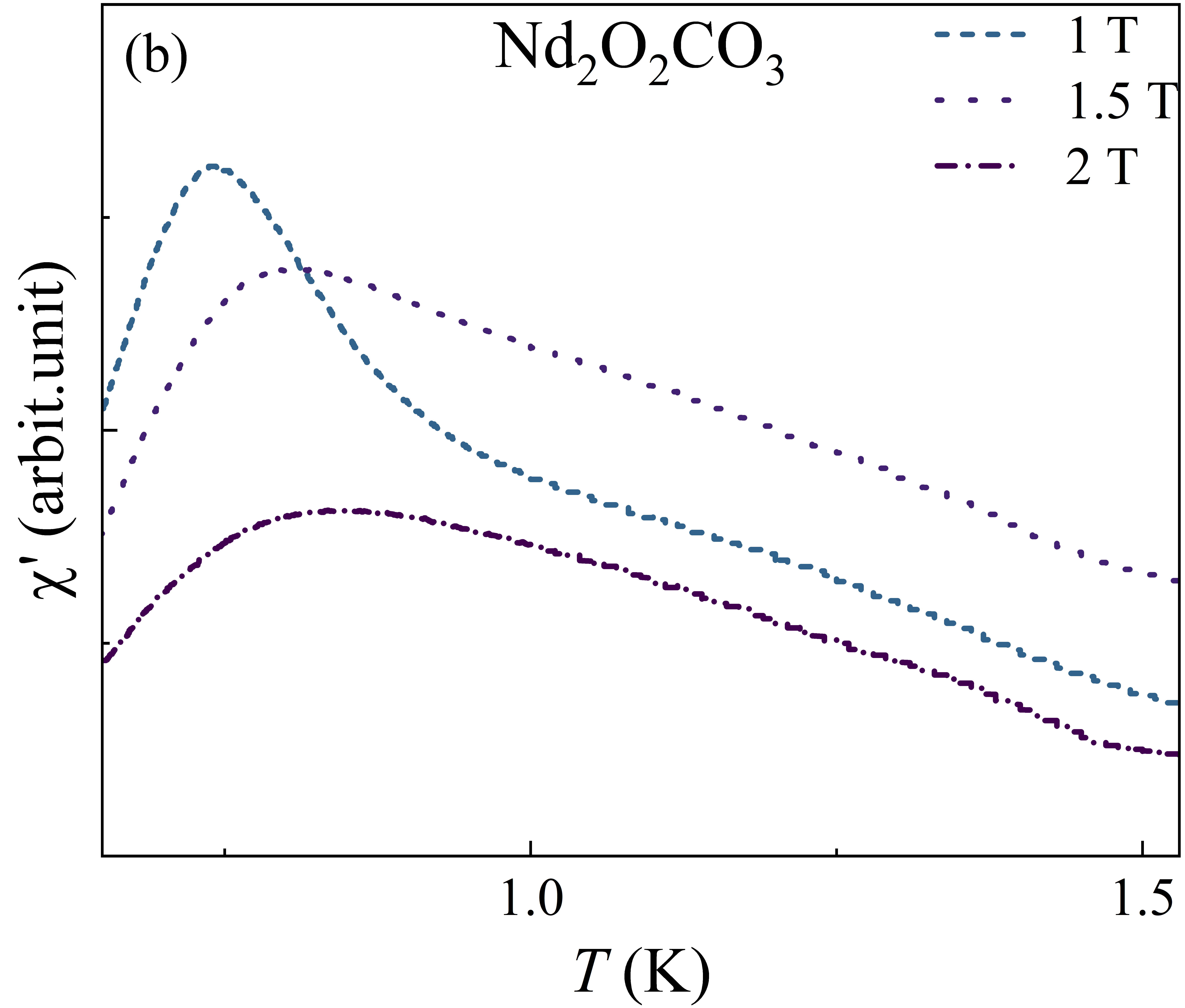}
        \\
        \includegraphics[width=0.67\columnwidth]{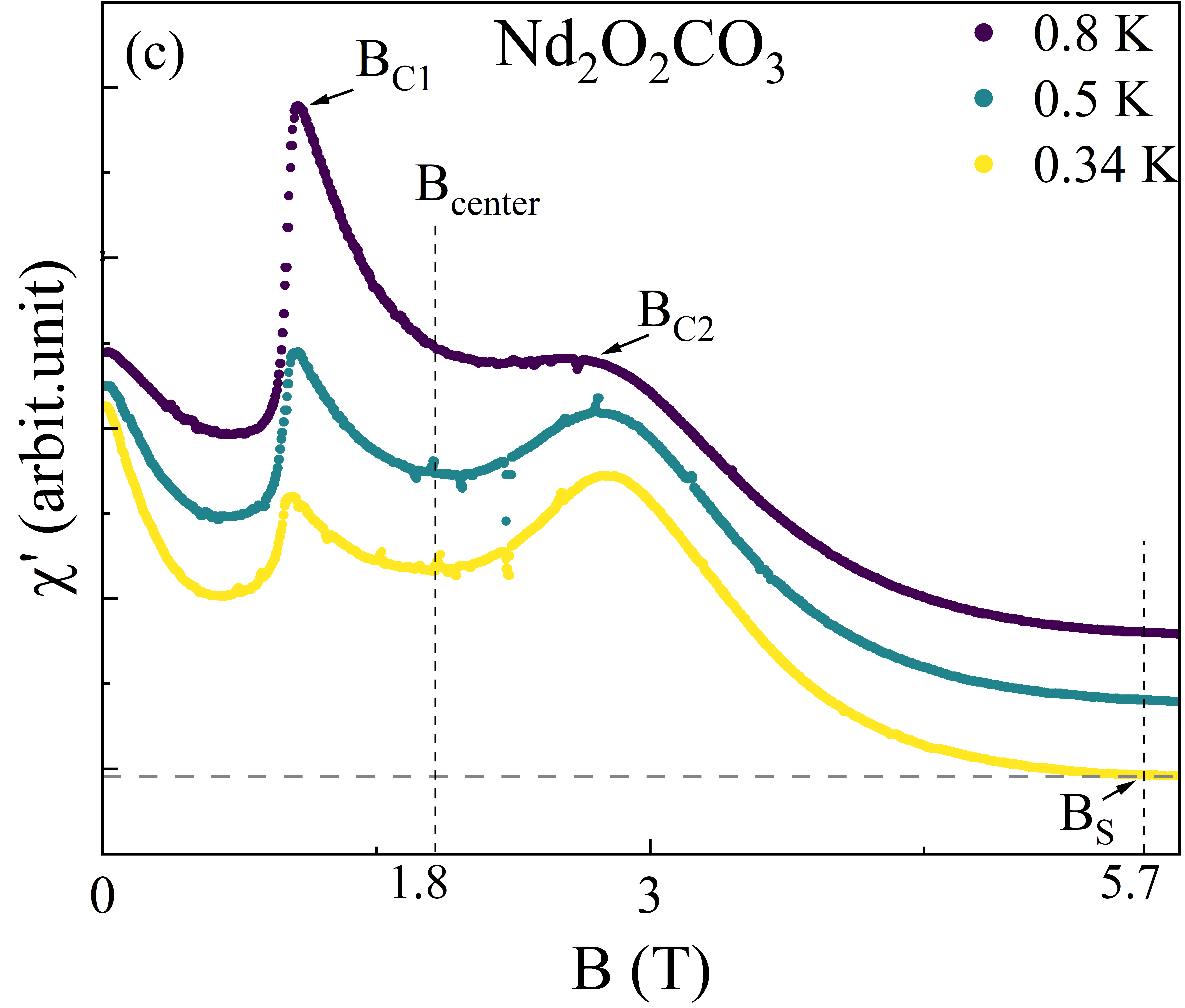}
        \includegraphics[width=0.66\columnwidth]{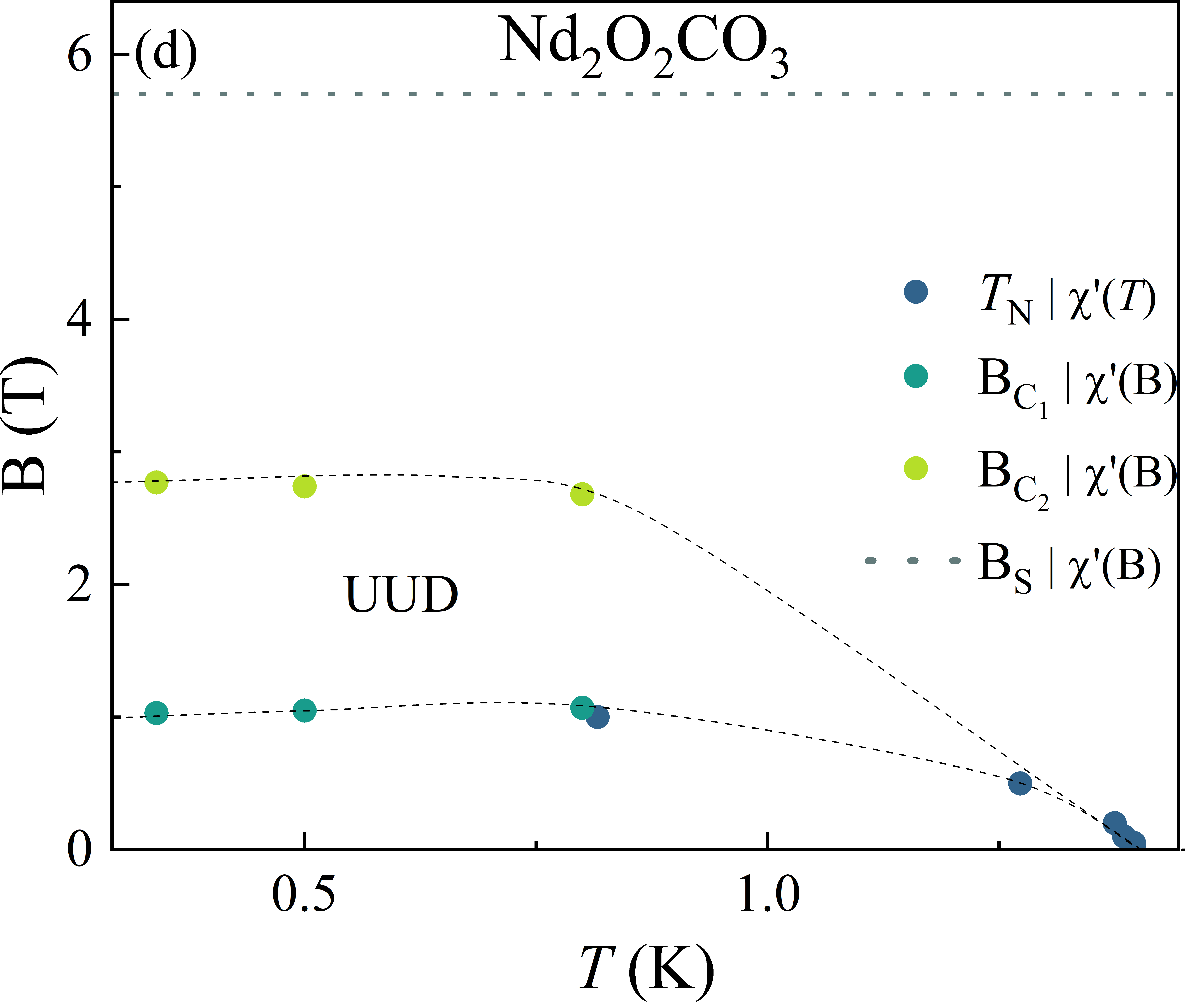}
        \caption{\ch{Nd_{2}O_{2}CO_{3}} (a): Temperature dependence of $\chi'$ measured at various low DC fields. The derivative of 0 T data is marked in red. (b): Temperature dependence of $\chi'$ at DC fields 1--2 T. (c): DC field dependence of $\chi'$ at low temperatures. (d): Magnetic phase diagram. The possible UUD phase is labeled.}
        \label{fig:nd}
    \end{figure*}
    
    \subsection{\ch{Nd_{2}O_{2}CO_{3}}}
            
        For \ch{Nd_{2}O_{2}CO_{3}}, the high temperature CW fit of \sfrac{1}{$\chi$} (Fig. \ref{fig:sus}(b)) yields a $\theta_{\text{CW}}$ = -48.51 K and a $\mu_{\text{eff}}$ = 3.79 $\mu_{\text{B}}$. This $\mu_{\text{eff}}$ value is consistent with the free-ion moment of $\mu_{\text{eff}}^{\text{theo}}$ = 3.62 $\mu_{\text{B}}$ expected for \ch{Nd^{3+}} ions. The low temperature CW fit yields a $\theta_{\text{CW}}$ = -4.13 K and a $\mu_{\text{eff}}$ = 2.66 $\mu_{\text{B}}$. 

        For \ch{Nd_{2}O_{2}CO_{3}}, a fast drop of its AC susceptibility (Fig. \ref{fig:nd}(a)), $\chi'(T)$, or a sharp peak of d$\chi'$/d$T$, occurs around 1.37 K, following a broad peak around 1.5 K. Such a feature suggests emerging long range magnetic order at $T_{\text{N}}$ = 1.37 K. With increasing DC magnetic field, the position of the fast drop shifts towards lower temperatures and the peak becomes even broader. Meanwhile, the DC field scan of $\chi'(\text{B})$ (Fig. \ref{fig:nd}(c)) measured at base temperature $T$ = 0.34 K shows two peaks at \ch{B_{C1}} = 1.04 T and \ch{B_{C2}} = 2.75 T and then becomes flat around \ch{B_{S}} = 5.7 T. In principal, $\chi'(\text{B})$ measures the derivative of the DC magnetization M(B), therefore the peaks of $\chi'(\text{B})$ suggest the existence of field induced spin state transitions and the flatness at high field suggests the emergence of a spin-polarized state. Accordingly, \ch{B_{S}} is the saturation field. With increasing temperature, these two peak positions of \ch{B_{C1}} and \ch{B_{C2}} exhibit little change. Using the critical temperatures and critical fields, a magnetic phase diagram was constructed in Fig. \ref{fig:nd}(d).
        
    \begin{figure*}[htb!]
        \centering
        \includegraphics[width=0.67\columnwidth]{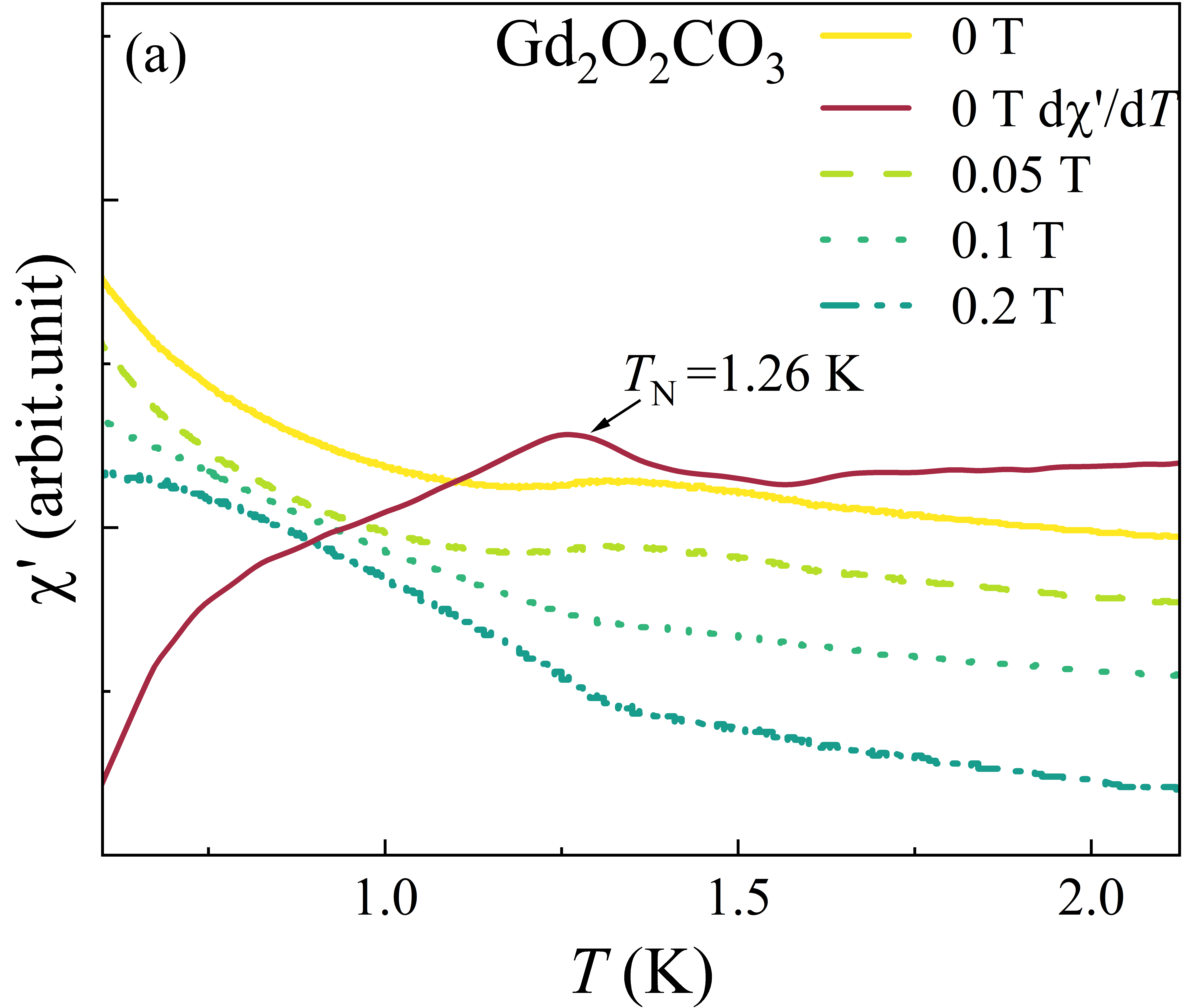}
        \includegraphics[width=0.67\columnwidth]{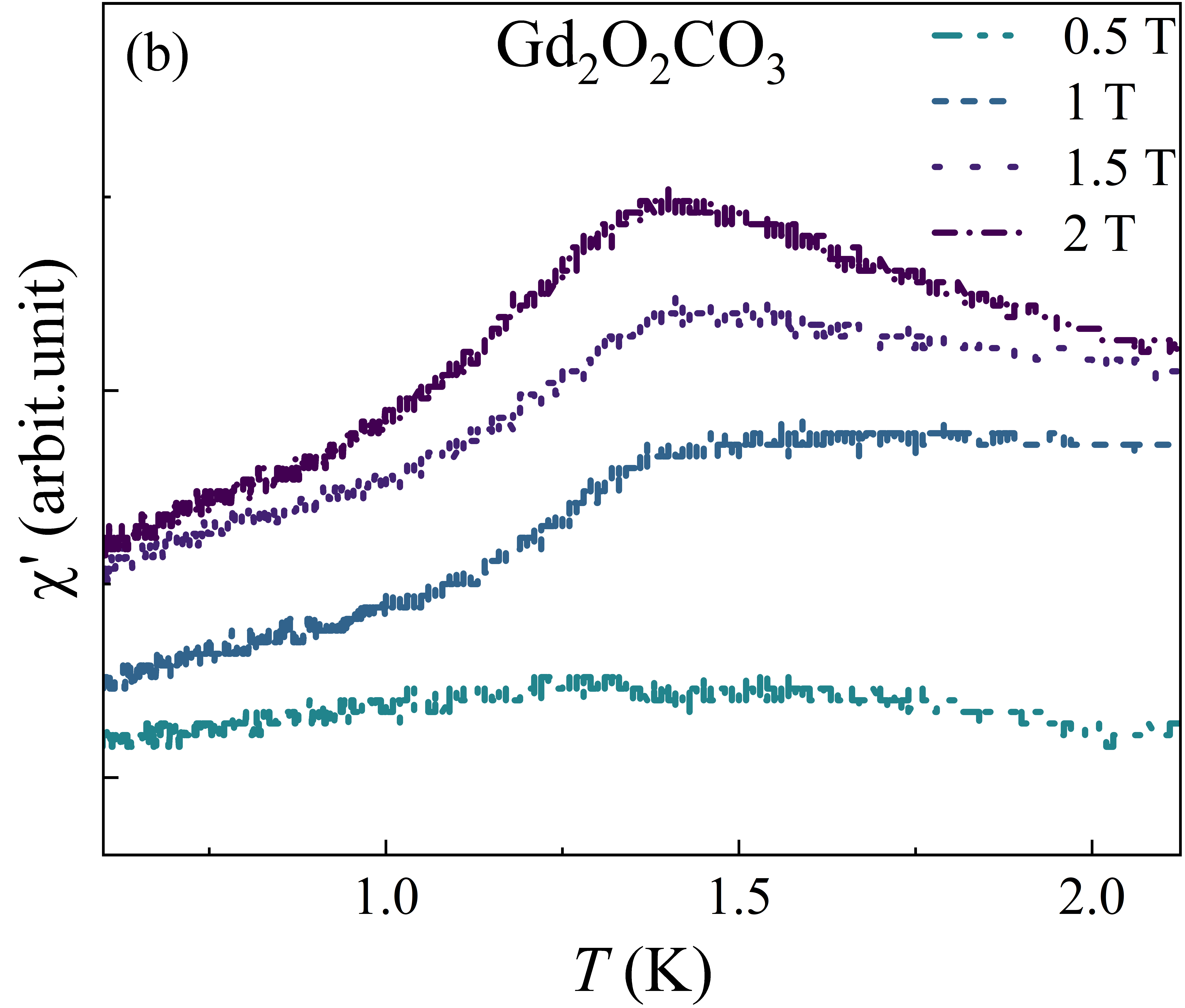}
        \includegraphics[width=0.67\columnwidth]{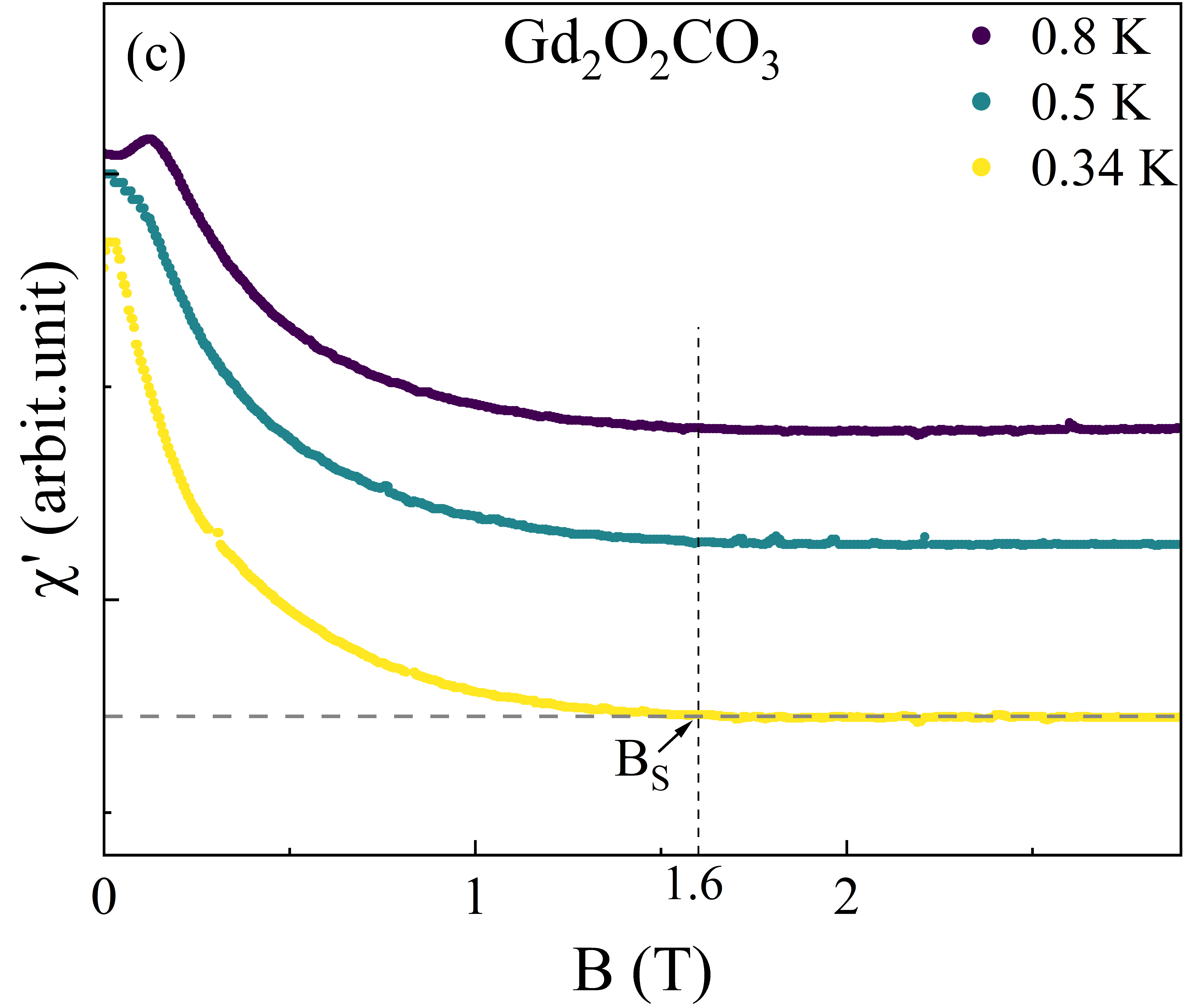}
        \caption{\ch{Gd_{2}O_{2}CO_{3}} (a): Temperature dependence of $\chi'$ measured at various low DC fields. The derivative of 0 T data is marked in red. (b):Temperature dependence of $\chi'$ at higher DC fields. (c): DC field dependence of $\chi'$ at low temperatures.}
        \label{fig:gd}
    \end{figure*}
    
    \subsection{\ch{Gd_{2}O_{2}CO_{3}}}
    
        For \ch{Gd_{2}O_{2}CO_{3}}, the CW fit of \sfrac{1}{$\chi$} from 150 to 300 K (Fig. \ref{fig:sus}(c)) yields a $\theta_{\text{CW}}$ = -5.42 K and a $\mu_{\text{eff}}$ = 8.97 $\mu_{\text{B}}$. This $\mu_{\text{eff}}$ value is consistent with the free-ion moment of $\mu_{\text{eff}}^{\text{theo}}$ = 7.94 $\mu_{\text{B}}$ expected for \ch{Gd^{3+}} ions. The $\chi'(T)$ (Fig. \ref{fig:gd}(a)) shows a drop, or a peak of d$\chi'$/d$T$, around $T_{\text{N}}$ = 1.26 K, which suggests long range magnetic ordering. With increasing DC magnetic field, this feature becomes weaker and a broad peak develops. The $\chi'(\text{B})$ data (Fig. \ref{fig:gd}(c)) suggests a saturation field around 1.6 T.
        
    \begin{figure*}[htb!]
        \centering
        \includegraphics[width=0.67\columnwidth]{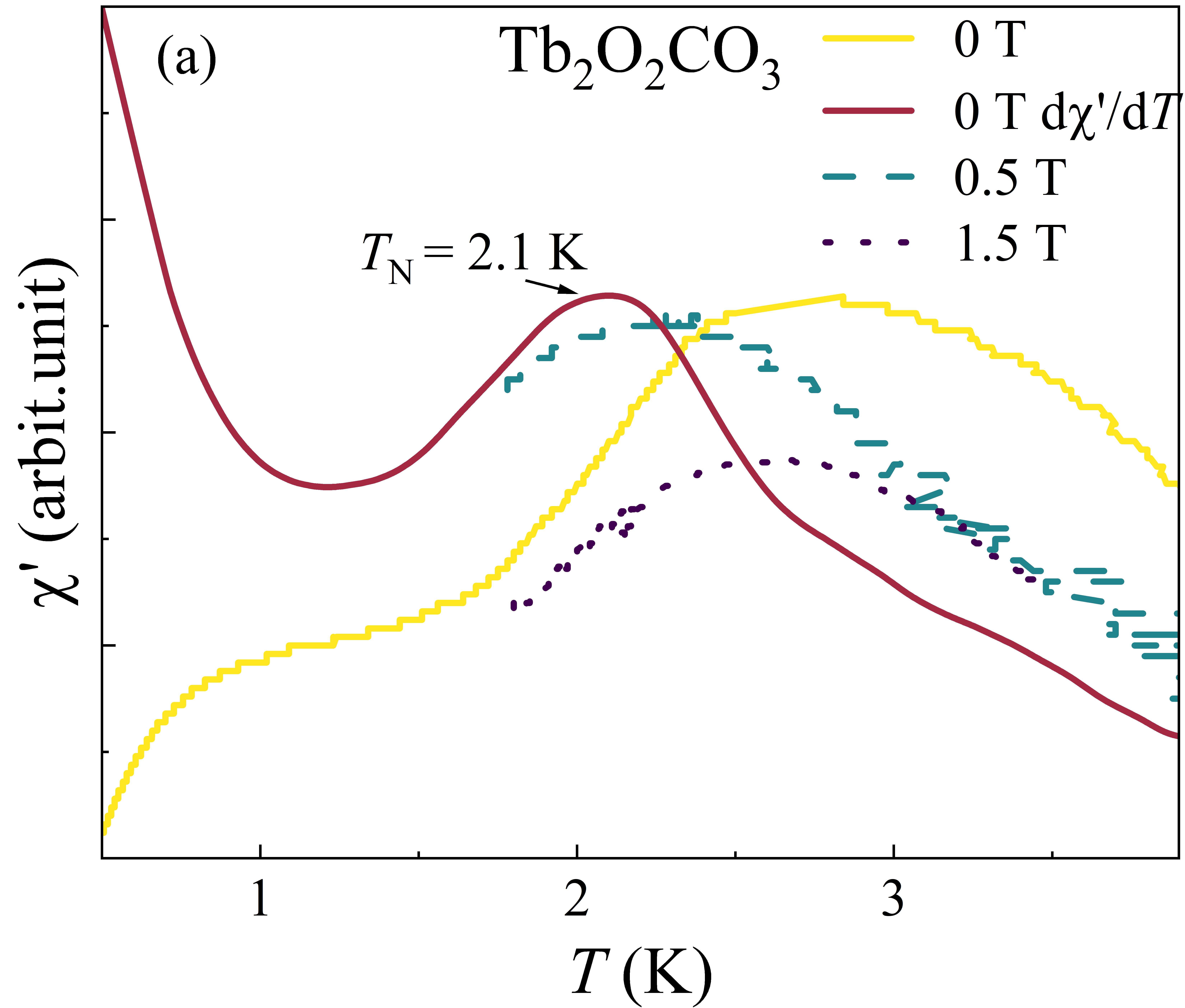}
        \includegraphics[width=0.67\columnwidth]{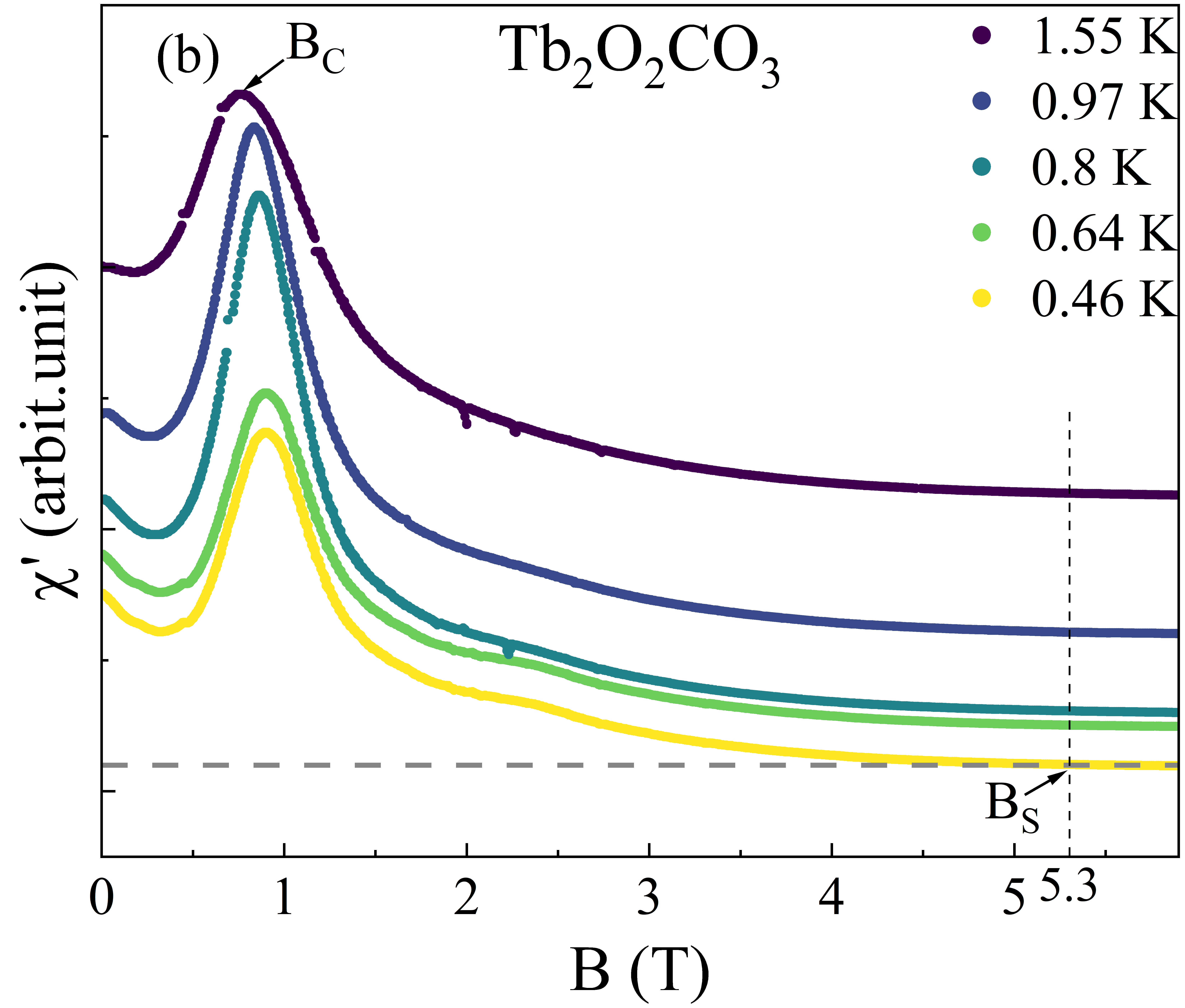}
        \includegraphics[width=0.67\columnwidth]{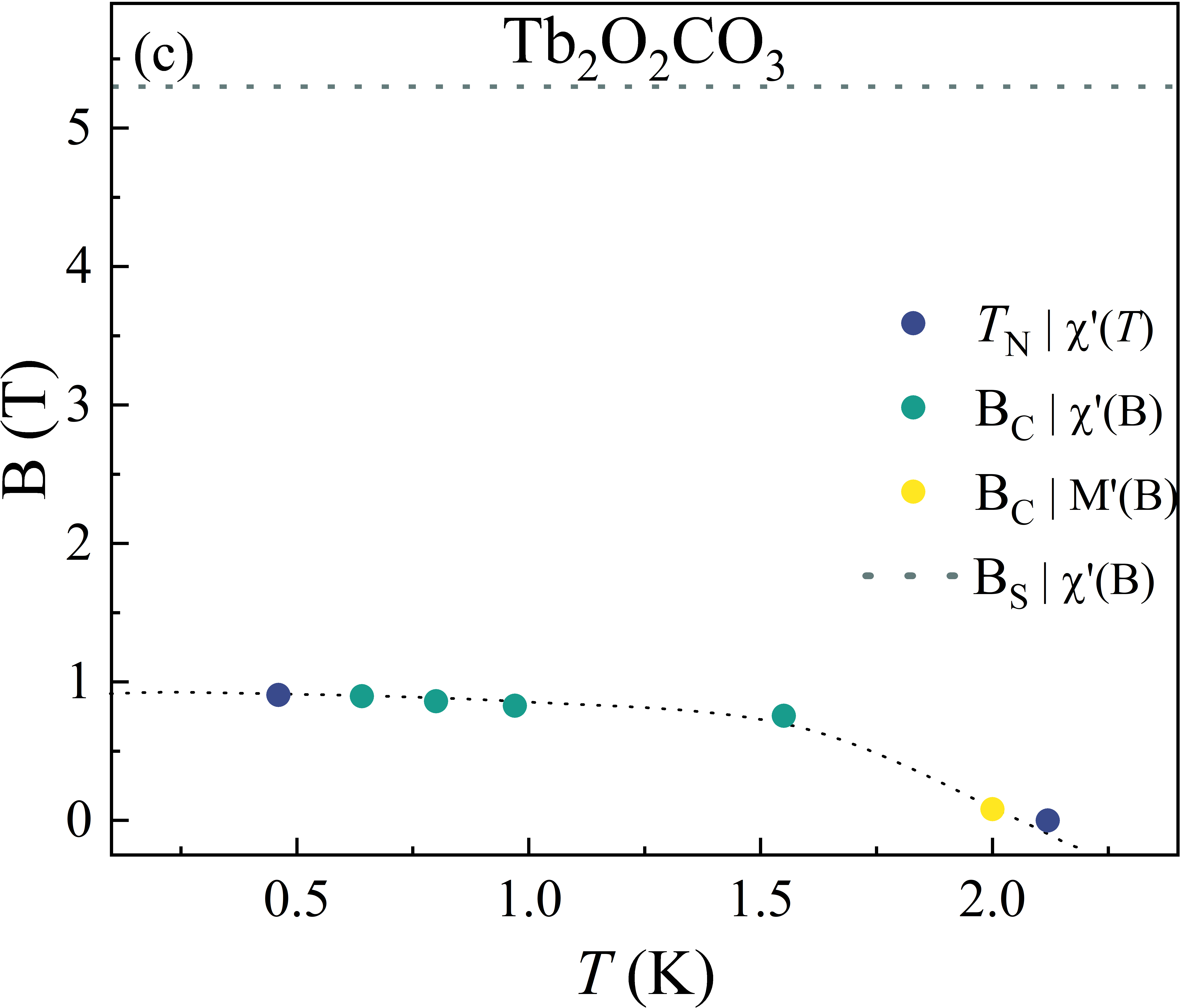}
        \caption{\ch{Tb_{2}O_{2}CO_{3}} (a): Temperature dependence of $\chi'$ measured at various DC fields. The derivative of 0 T data is marked in red.  (b): DC field dependence of $\chi'$ at low temperatures. (c): Magnetic phase diagram.}
        \label{fig:tb}
    \end{figure*}
    
    \subsection{\ch{Tb_{2}O_{2}CO_{3}}}
    
       For \ch{Tb_{2}O_{2}CO_{3}}, the high temperature CW fit of \sfrac{1}{$\chi$} (Fig. \ref{fig:sus}(d)) yields a $\theta_{\text{CW}}$ = -9.06 K and a $\mu_{\text{eff}}$ = 9.52 $\mu_{\text{B}}$. This $\mu_{\text{eff}}$ value is consistent with the free-ion moment of $\mu_{\text{eff}}^{\text{theo}}$ = 9.72 $\mu_{\text{B}}$ expected for \ch{Tb^{3+}} ions. The low temperature CW fit yields a $\theta_{\text{CW}}$ = -6.70 K and a $\mu_{\text{eff}}$ = 8.85 $\mu_{\text{B}}$.
       
       For \ch{Tb_{2}O_{2}CO_{3}}, the $\chi'(T)$ (Fig. \ref{fig:tb}(a)) shows a drop, or a peak of d$\chi'$/d$T$, around 2.10 K, following a broad peak around 2.8 K. This suggests long range magnetic ordering at $T_{\text{N}}$ = 2.10 K. With increasing DC magnetic field, this features shifts towards lower temperatures. The $\chi'(\text{B})$ data (Fig. \ref{fig:tb}(b)) measured at base temperature $T$ = 0.46 K shows a peak at \ch{B_{C}} = 0.90 T and becomes flat around \ch{B_{S}} = 5.3 T. Correspondingly, the derivative of DC magnetization, M(B), measured at 2.0 K also shows a peak at  \ch{B_{C}} = 0.8 T, as plotted in Fig. \ref{fig:mag}(d). With increasing temperature, this peak position remains relatively the same. A magnetic phase diagram was constructed in Fig. \ref{fig:tb}(c).

    \begin{figure*}[htb!]
        \centering
        \includegraphics[width=0.67\columnwidth]{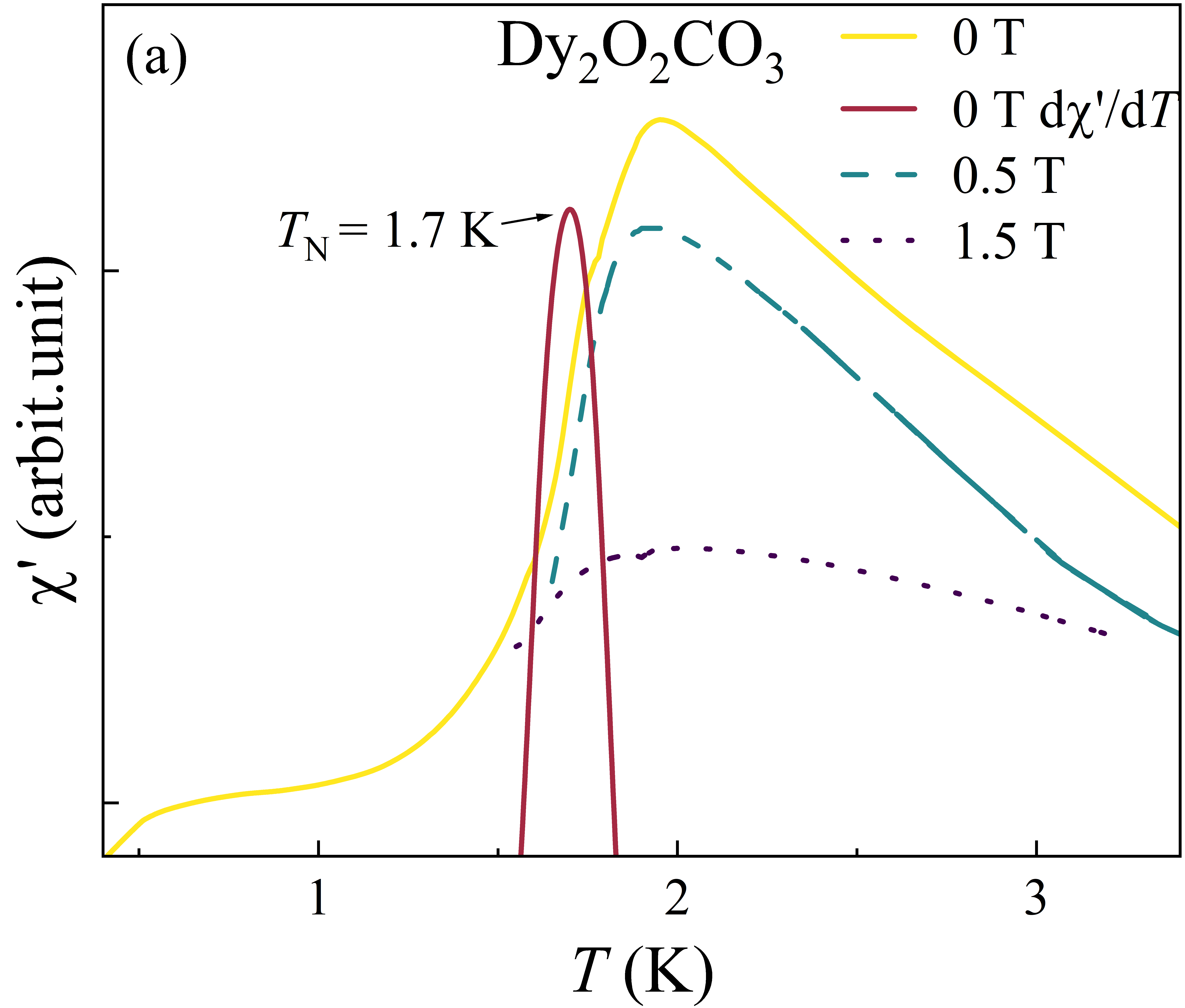}
        \includegraphics[width=0.67\columnwidth]{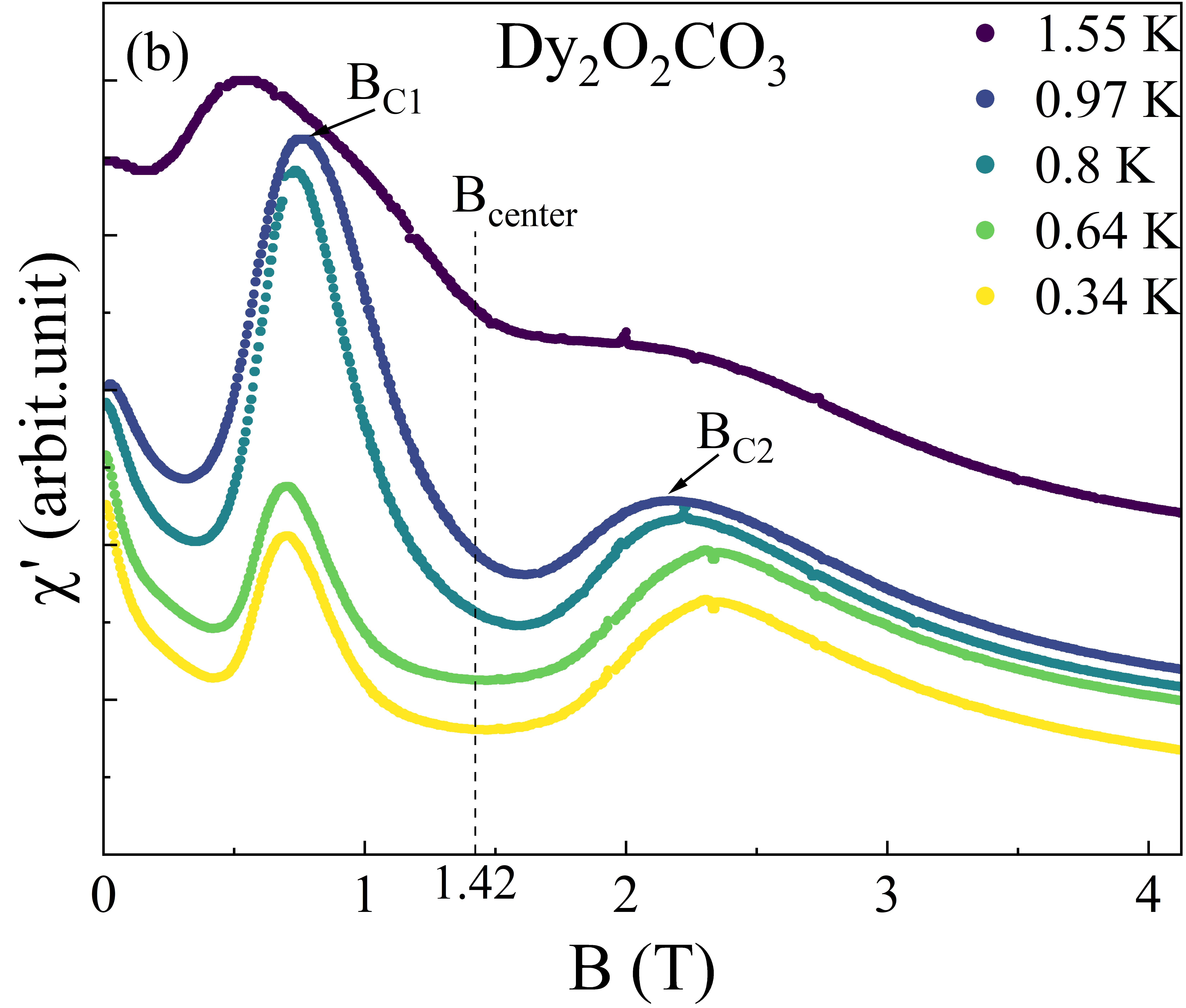}
        \includegraphics[width=0.67\columnwidth]{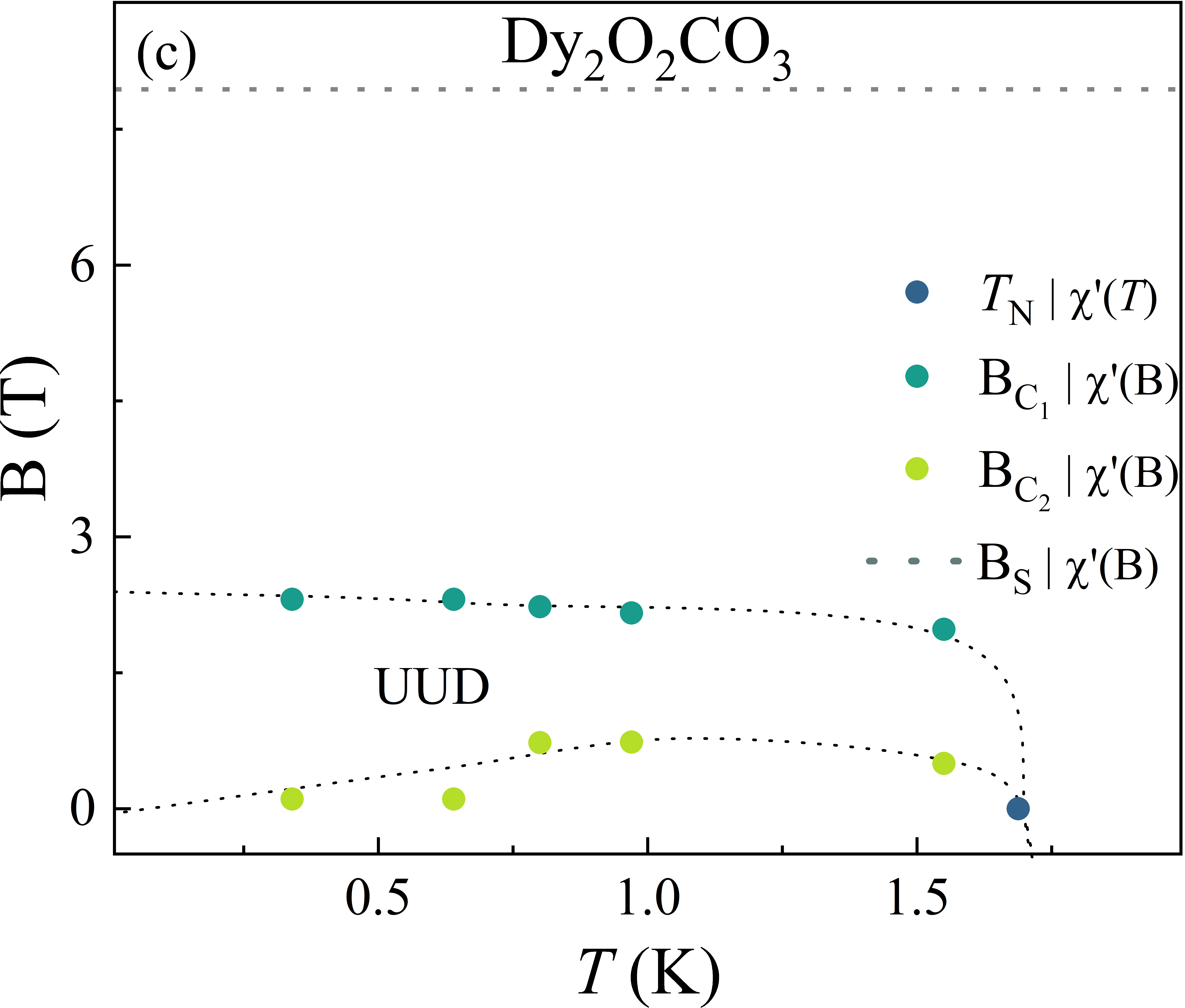}
        \caption{\ch{Dy_{2}O_{2}CO_{3}} (a): Temperature dependence of $\chi'$ measured at various DC fields. The derivative of 0 T data is marked in red.  (b): DC field dependence of $\chi'$ at low temperatures. (c): Magnetic phase diagram. The possible UUD phase is labeled.}
        \label{fig:dy}
    \end{figure*}
    
    \subsection{\ch{Dy_{2}O_{2}CO_{3}}}
    
        For \ch{Dy_{2}O_{2}CO_{3}}, the high temperature CW fit of \sfrac{1}{$\chi$} (Fig. \ref{fig:sus}(e)) yields a $\theta_{\text{CW}}$ = -1.30 K and a $\mu_{\text{eff}}$ = 10.20 $\mu_{\text{B}}$. This $\mu_{\text{eff}}$ value is consistent with the free-ion moment of $\mu_{\text{eff}}^{\text{theo}}$ = 10.65 $\mu_{\text{B}}$ expected for \ch{Dy^{3+}} ions. The low temperature CW fit yields a $\theta_{\text{CW}}$ = -8.70 K and a $\mu_{\text{eff}}$ = 9.33 $\mu_{\text{B}}$. 

        For \ch{Dy_{2}O_{2}CO_{3}}, a fast drop of $\chi'(T)$ (Fig. \ref{fig:dy}(a)), or a sharp peak of d$\chi'$/d$T$, occurs around 1.70 K, following a broad peak around 2.0 K. This suggests long range magnetic ordering at $T_{\text{N}}$ = 1.70 K. With increasing DC magnetic field, the fast drop position shifts to lower temperatures and peak becomes even broader. Meanwhile, $\chi'(\text{B})$ (Fig. \ref{fig:dy}(b)) measured at base temperature $T$ = 0.34 K shows two peaks at \ch{B_{C1}} = 0.70 T and \ch{B_{C2}} = 2.32 T, and then becomes flat around \ch{B_{S}} = 9 T (not shown here). With increasing temperature, these two peak positions change little, until $T$ = 1.55 K, wherein the peaks significantly broaden and shift toward lower fields. Its magnetic phase diagram was constructed in Fig. \ref{fig:dy}(c).

    \begin{figure*}[htb!]
        \centering
        \includegraphics[width=0.67\columnwidth]{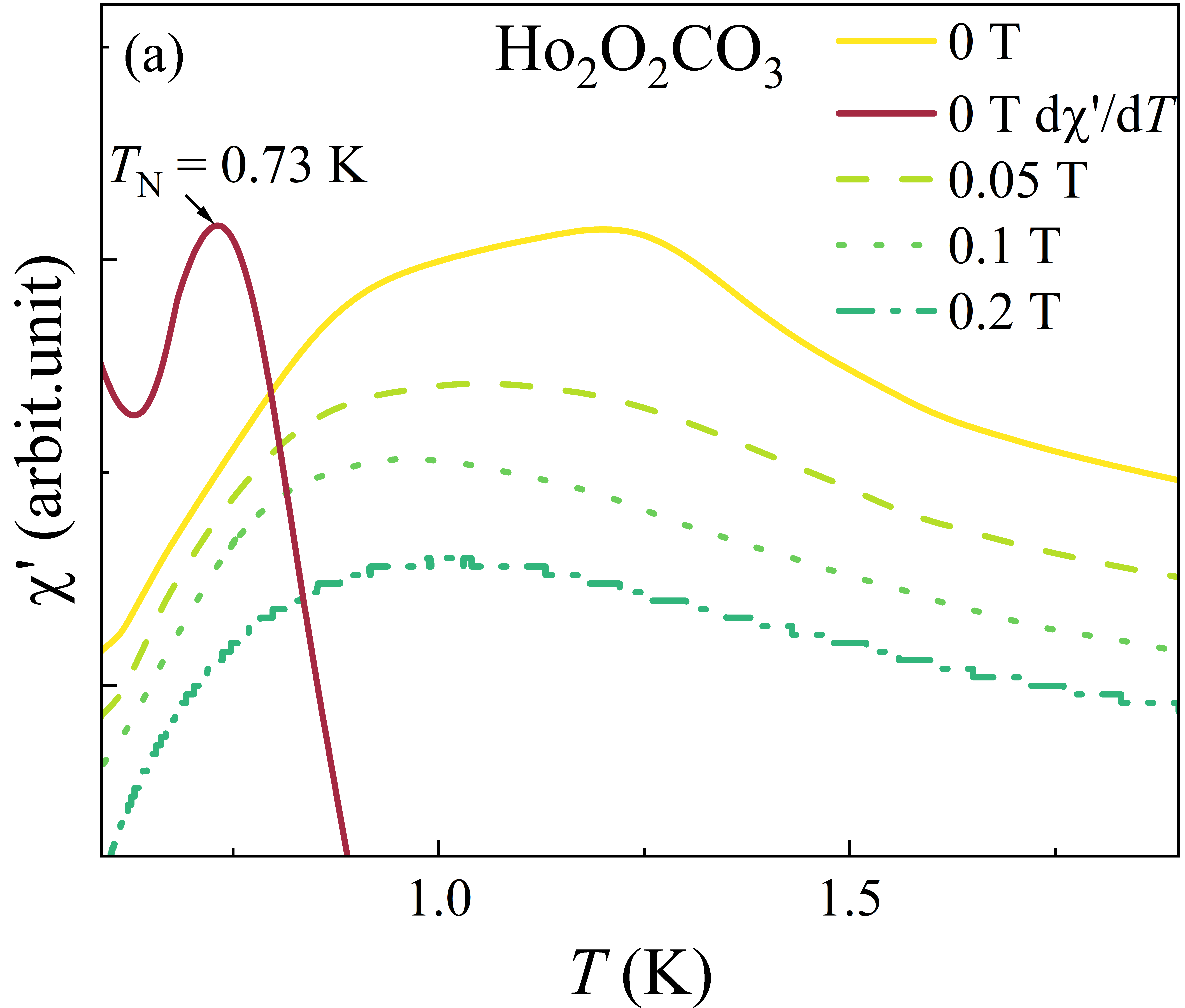}
        \includegraphics[width=0.67\columnwidth]{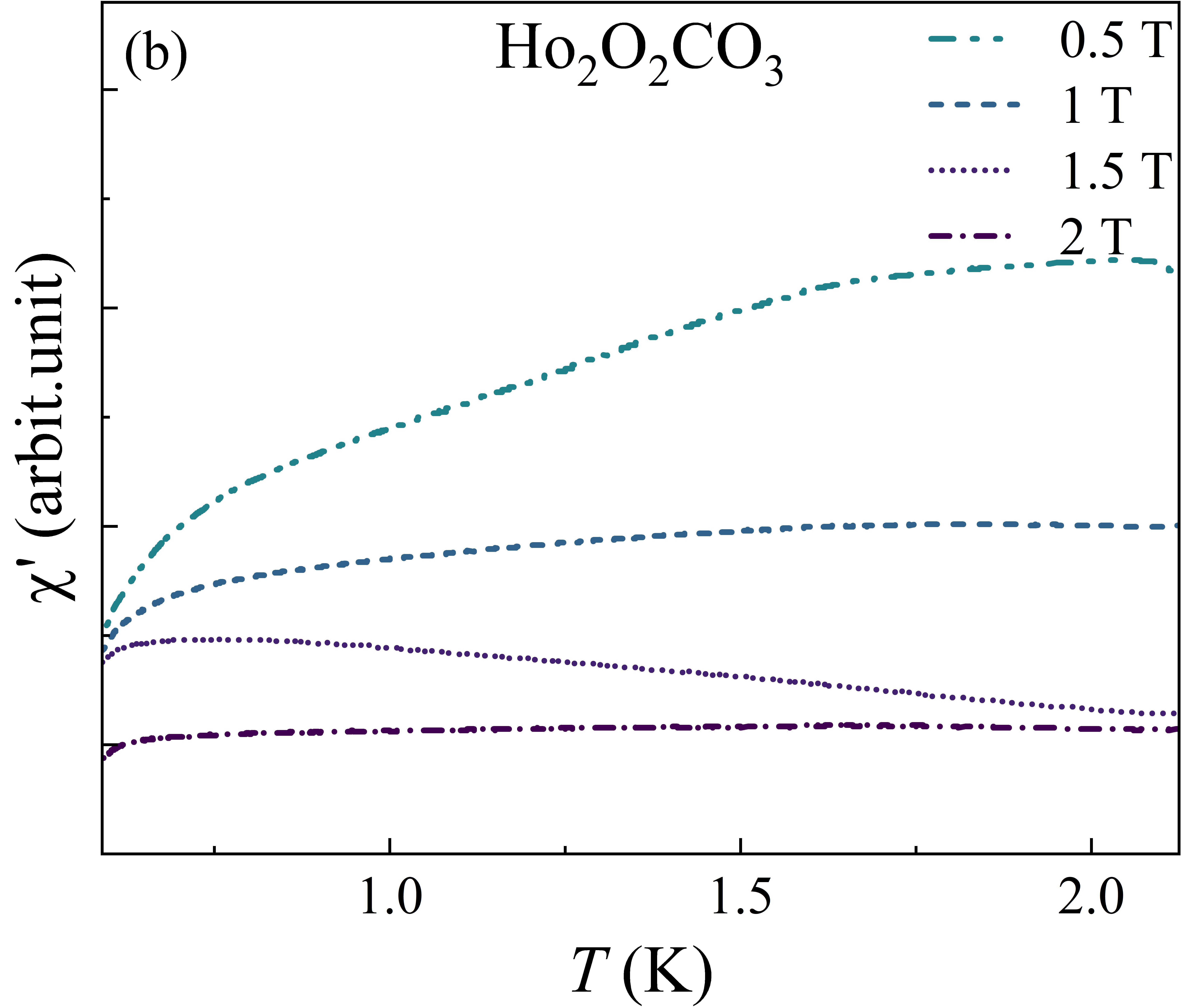}
        \\
        \includegraphics[width=0.67\columnwidth]{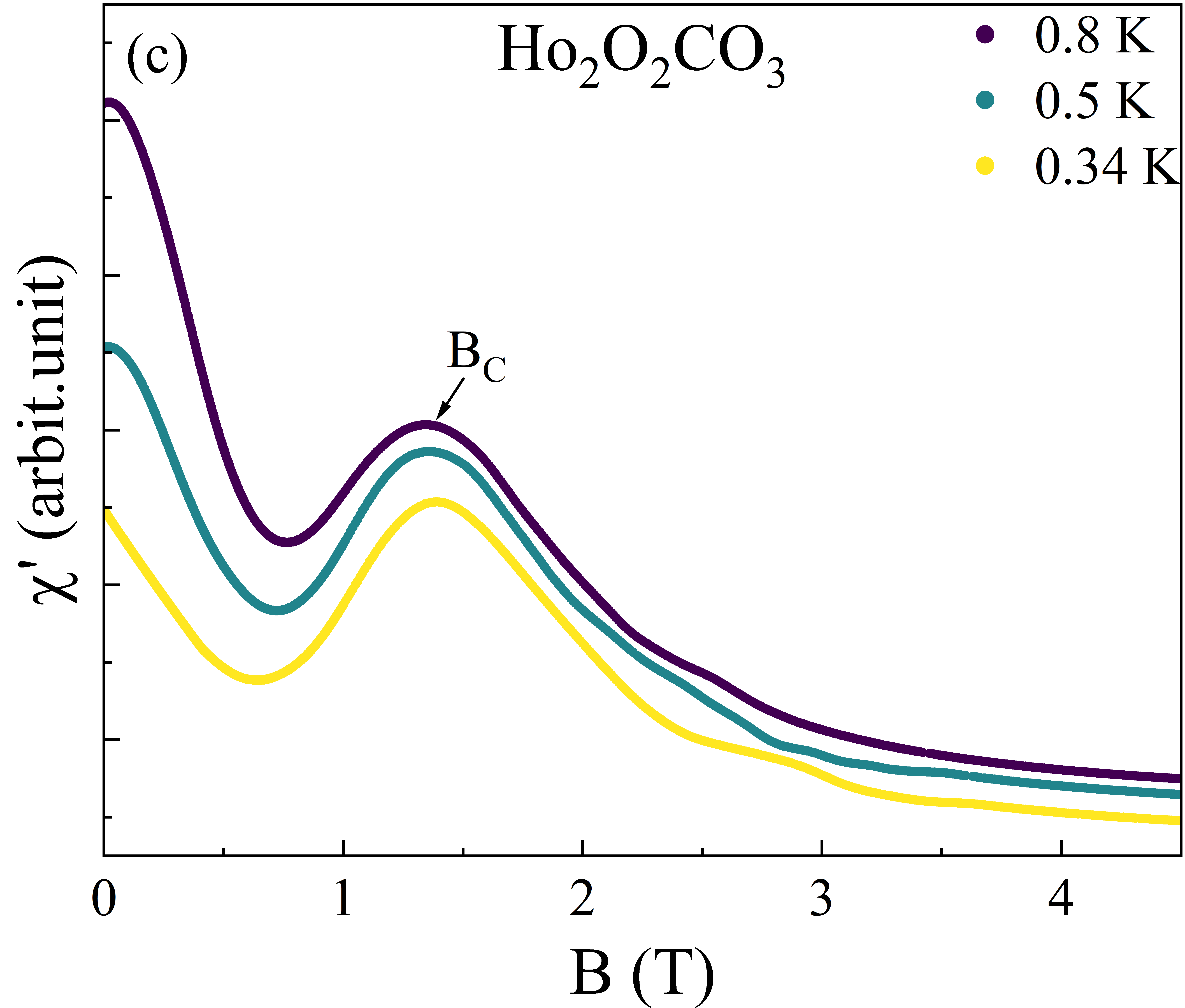}
        \includegraphics[width=0.67\columnwidth]{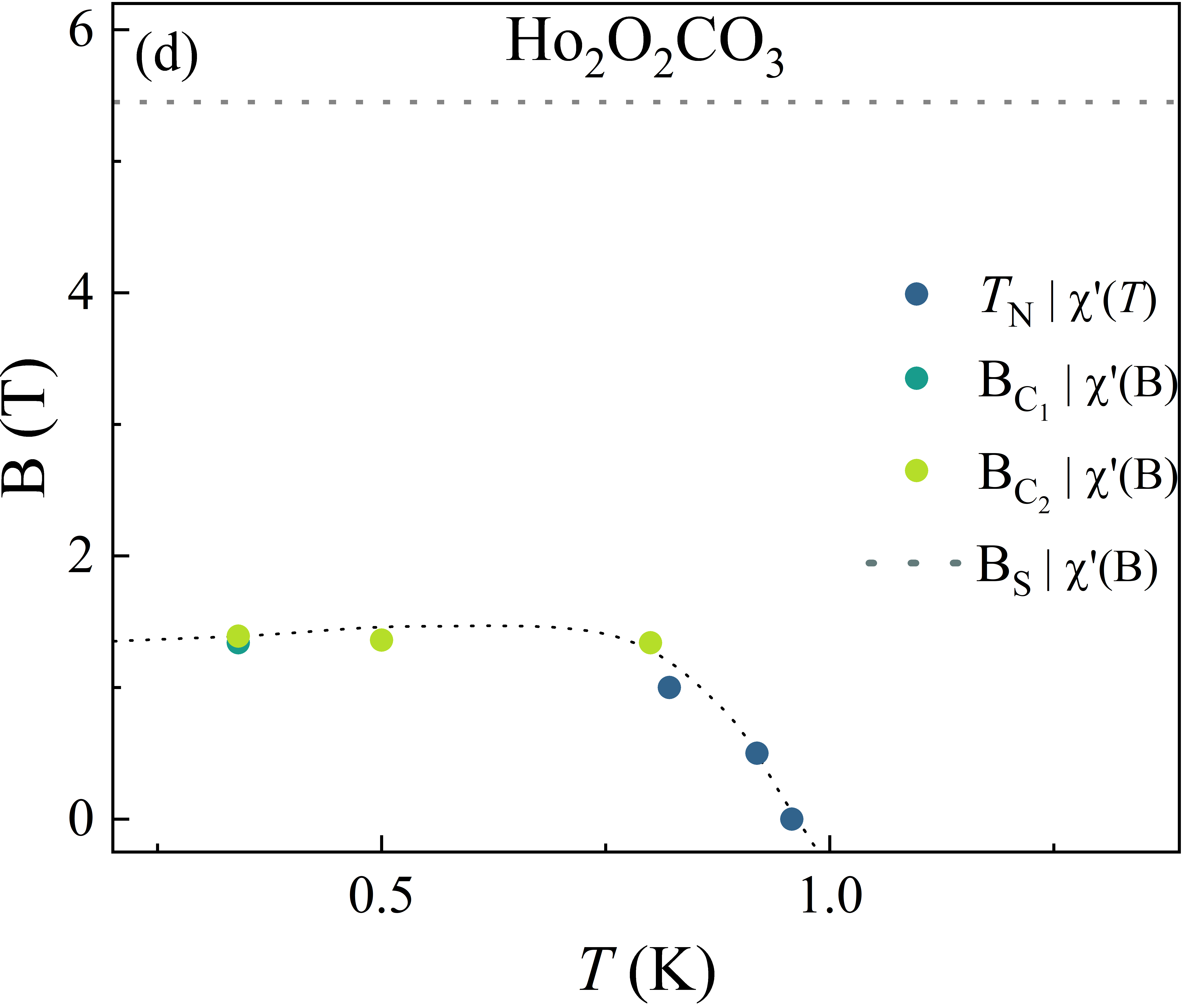}
        \caption{\ch{Ho_{2}O_{2}CO_{3}} (a): Temperature dependence of $\chi'$ measured at various low DC fields. The derivative of 0 T data is marked in red. (b): Temperature dependence of $\chi'$ at DC fields 0.5--2 T. (c): DC field dependence of $\chi'$ at low temperatures. (d): Magnetic phase diagram.}
        \label{fig:ho}
    \end{figure*}
    
    \subsection{\ch{Ho_{2}O_{2}CO_{3}}}
    
        For \ch{Ho_{2}O_{2}CO_{3}}, the high temperature CW fit of \sfrac{1}{$\chi$} (Fig. \ref{fig:mag}(f)) yields a $\theta_{\text{CW}}$ = -0.19 K and a $\mu_{\text{eff}}$ = 9.67 $\mu_{\text{B}}$. This $\mu_{\text{eff}}$ value is consistent with the free-ion moment of $\mu_{\text{eff}}^{\text{theo}}$ = 10.61 $\mu_{\text{B}}$ expected for \ch{Ho^{3+}} ions. The low temperature CW fit yields a $\theta_{\text{CW}}$ = -6.83 K and a $\mu_{\text{eff}}$ = 9.41 $\mu_{\text{B}}$. 
        
        $\chi'(T)$ (Fig. \ref{fig:ho}(a,b)) shows a drop, or a peak of d$\chi'$/d$T$, around 0.73 K, following a broad peak around 1.2 K. This suggests long range magnetic ordering at $T_{\text{N}}$ = 0.73 K. With increasing DC magnetic field, these features shift to low temperatures. The $\chi'(\text{B})$ data (Fig. \ref{fig:ho}(c)) measured at base temperature $T$ = 0.34 K shows a single peak at \ch{B_{C}} = 1.40 T and becomes flat around \ch{B_{S}} = 5.5 T (not shown here).  With increasing temperature, the peak position changes little. Its magnetic phase diagram was constructed Fig. \ref{fig:ho}(d).

    \begin{figure*}[htb!]
        \centering
        \includegraphics[width=0.67\columnwidth]{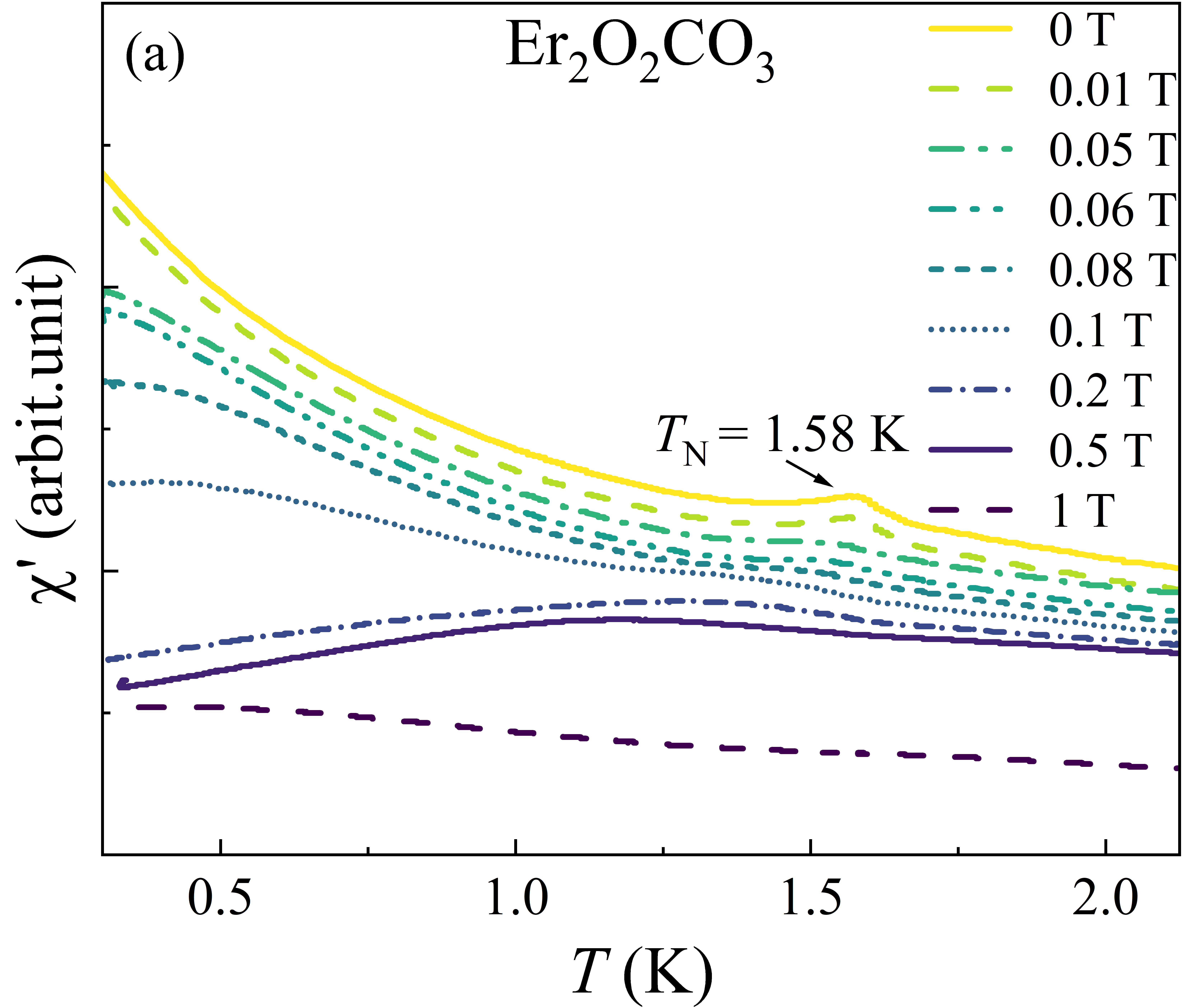}
        \includegraphics[width=0.67\columnwidth]{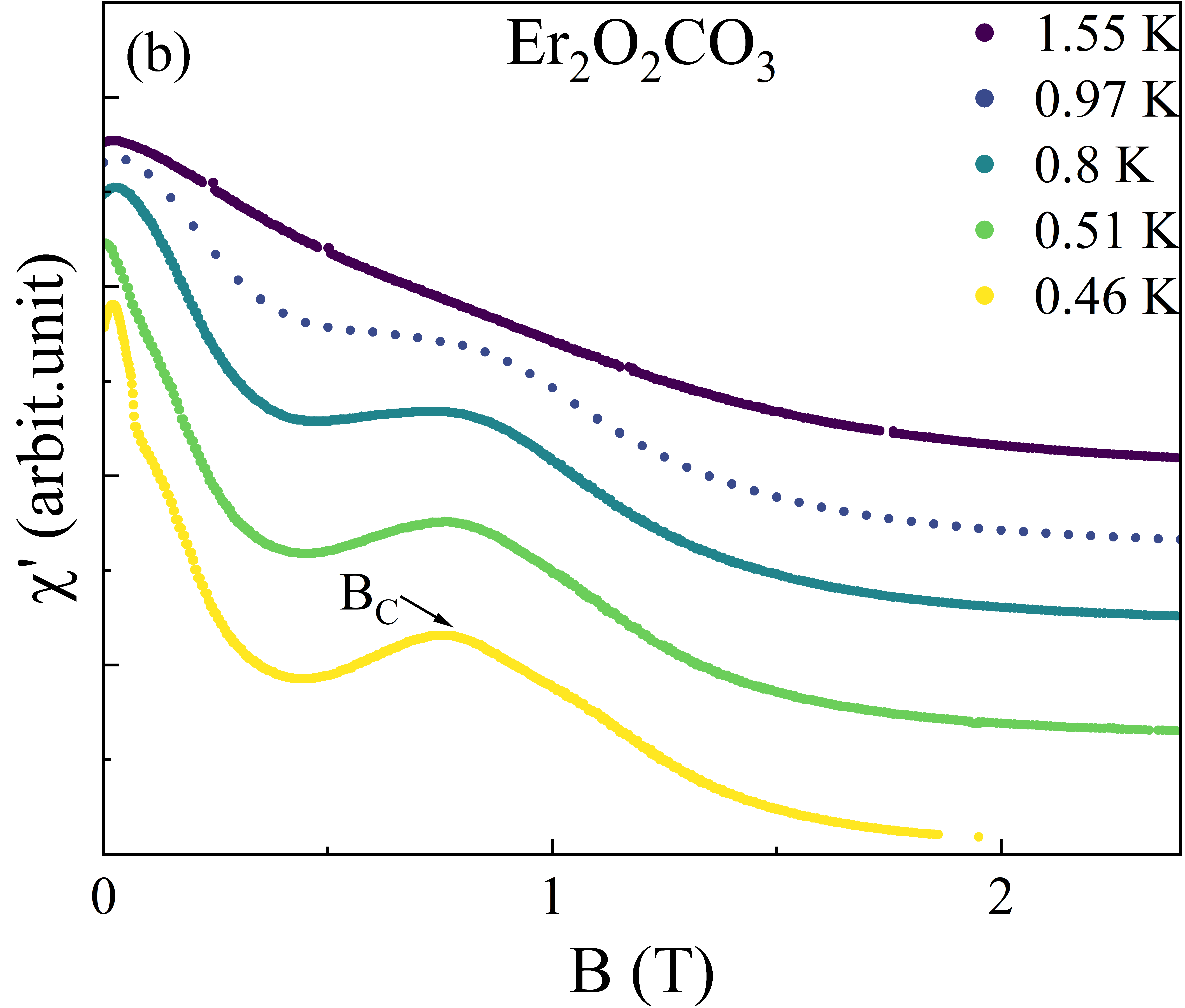}
        \includegraphics[width=0.67\columnwidth]{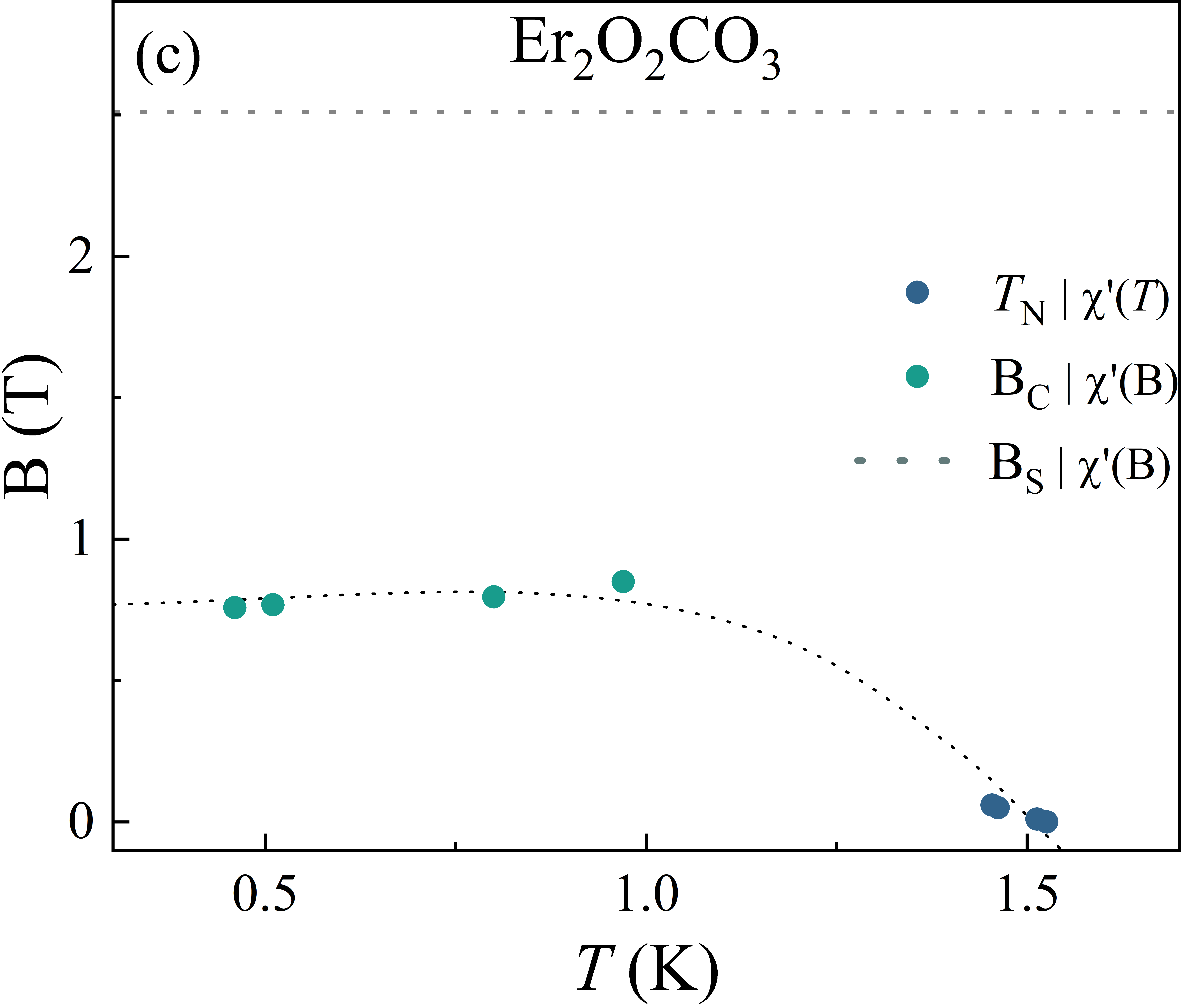}
        \caption{\ch{Er_{2}O_{2}CO_{3}} (a): Temperature dependence of $\chi'$ measured at various DC fields. (b): DC field dependence of $\chi'$ at low temperatures. (c): Magnetic phase diagram.}
        \label{fig:er}
    \end{figure*}

    \begin{figure}[htb!]
        \centering
        \includegraphics[width=0.85\columnwidth]{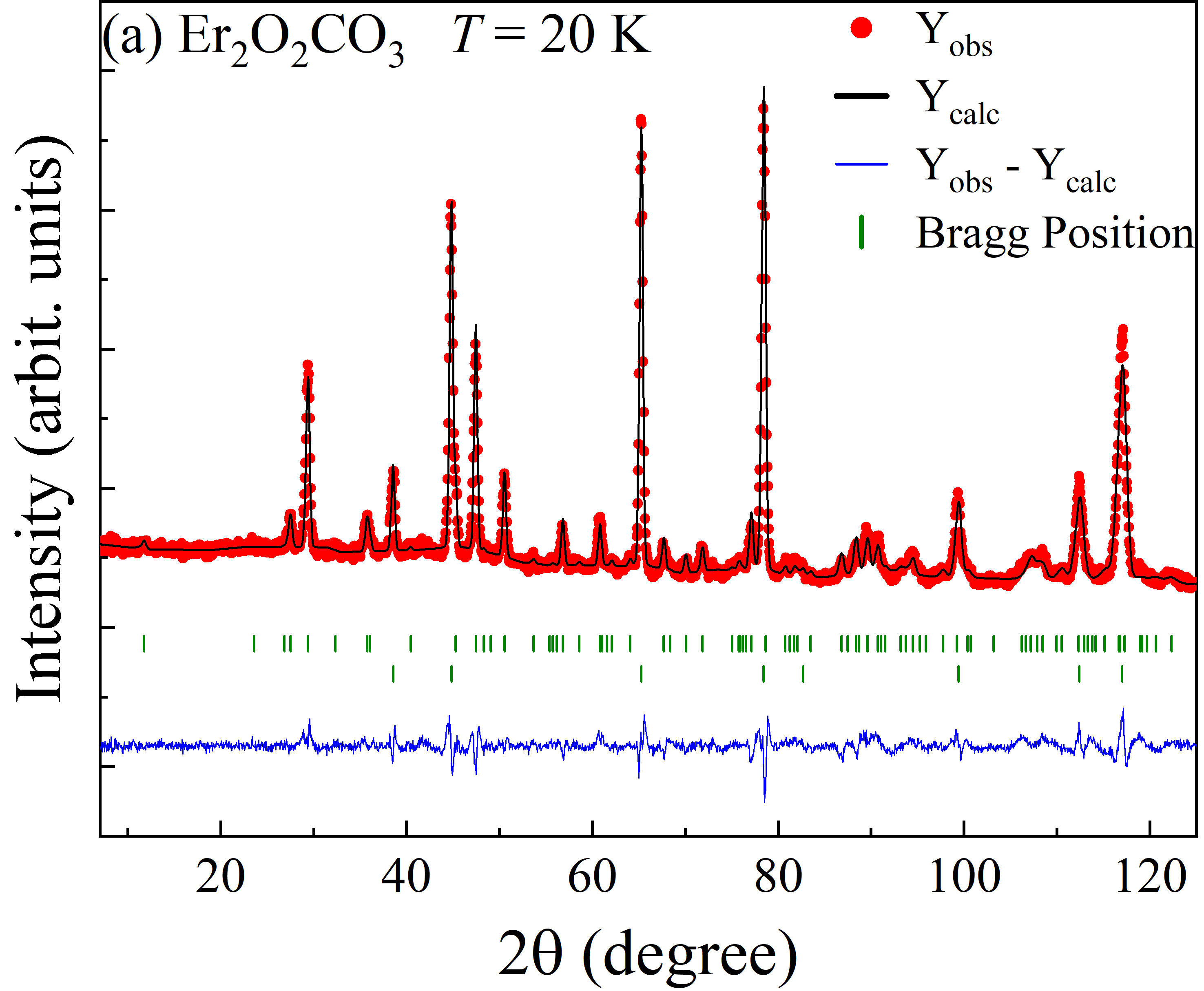}
        \includegraphics[width=0.85\columnwidth]{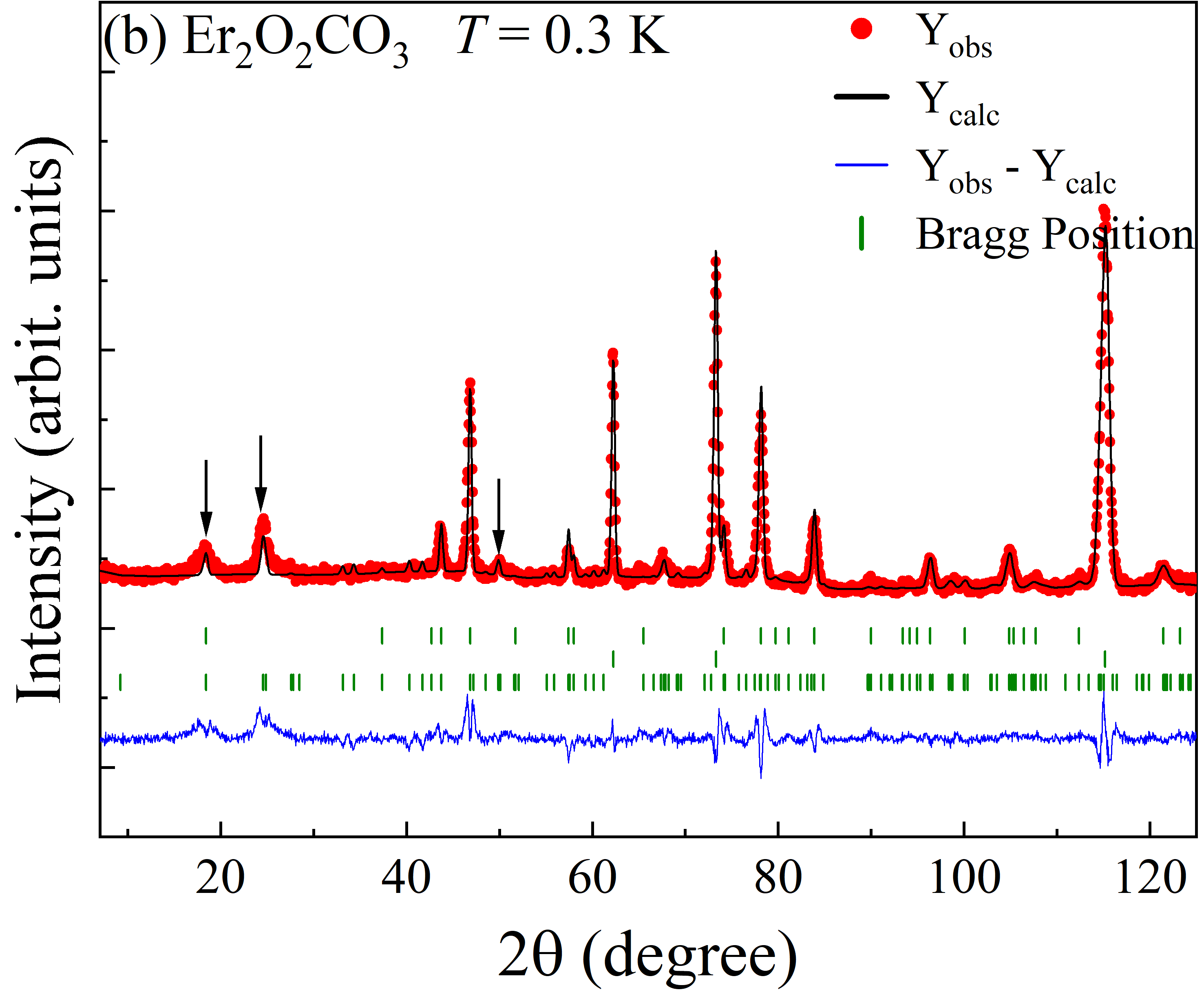}
        \caption{NPD data for \ch{Er_{2}O_{2}CO_{3}}. (a): The NPD pattern measured at 20 K with a neutron wavelength of 1.54\AA{} (red) and its Rietveld refinement. Green bars show the positions of nuclear Bragg peaks for \ch{Er_{2}O_{2}CO_{3}} (top) and the aluminum can (bottom) which held the sample during measurements. (b): The NPD data measured at 0.3 K with a neutron wavelength of 2.41\AA{} and its Rietveld refinement. The green bars show the positions of nuclear Bragg peaks of \ch{Er_{2}O_{2}CO_{3}} (top), Al can (middle), and magnetic Bragg peaks of \ch{Er_{2}O_{2}CO_{3}} (bottom). Black arrows highlight the main magnetic Bragg peaks.}
        \label{fig:ernpdr}
    \end{figure}

    \begin{figure}[htb!]
        \centering
        \includegraphics[width=1.0\columnwidth]{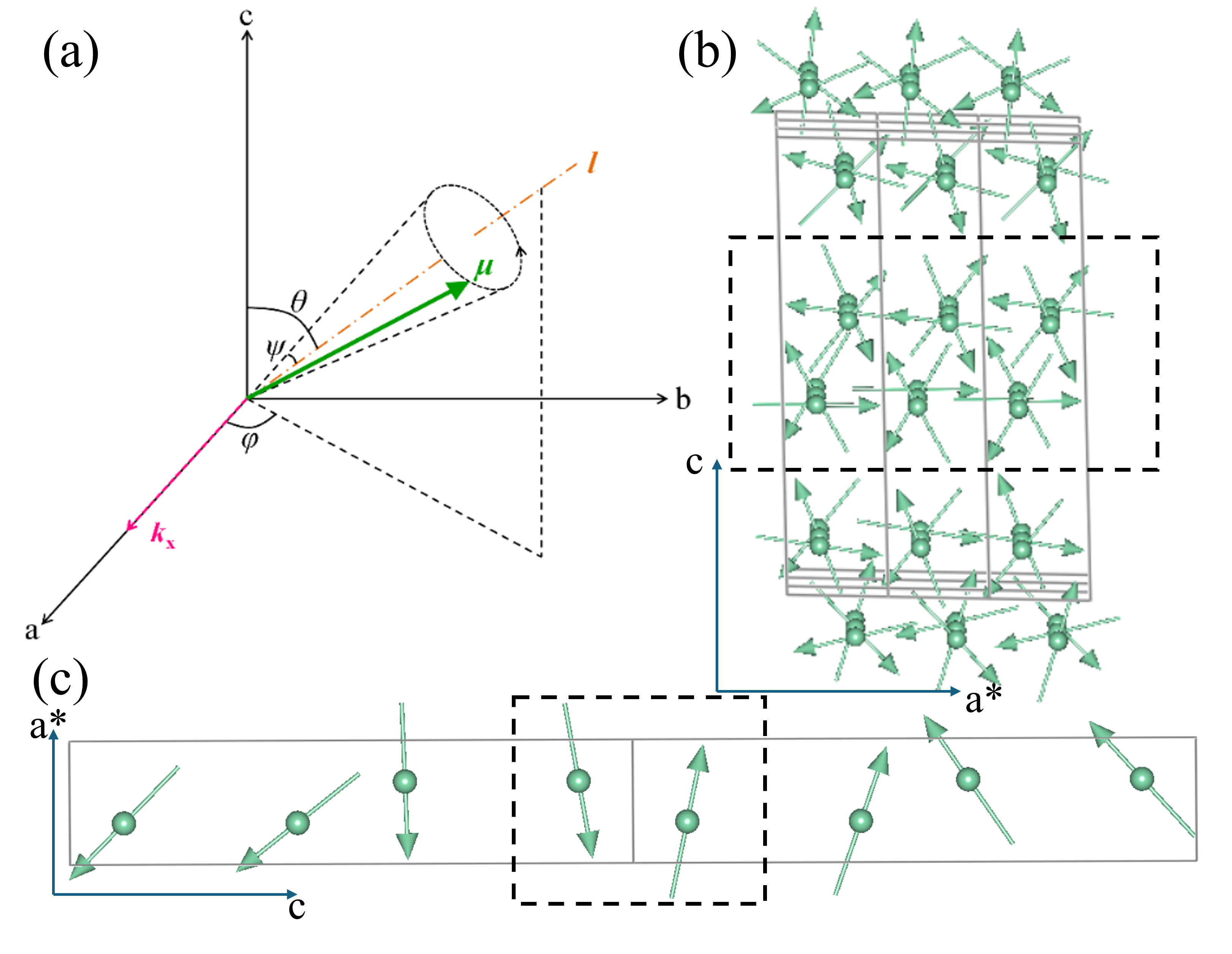}
        \caption{(a): A representation of spin as illustrated by the spiral model in the EDPCR software of the FULLPROF package. (b): A tentative spin structure of \ch{Er_{2}O_{2}CO_{3}} in its magnetic ordering state. All spins lie in the \textit{ac}-plane and form a 120-degree spin structure in the triangular plane (\textit{ab}-plane). (c): Meanwhile, each spin in a unit cell and its corresponding spins in other unit cells form a cycloid structure along the \textit{c}-axis. The black dashed rectangles highlight the bilayer triangular structure.}
        \label{fig:ernpdm}
    \end{figure}

    \begin{figure}[htb!]
        \centering
        \includegraphics[width=1.0\columnwidth]{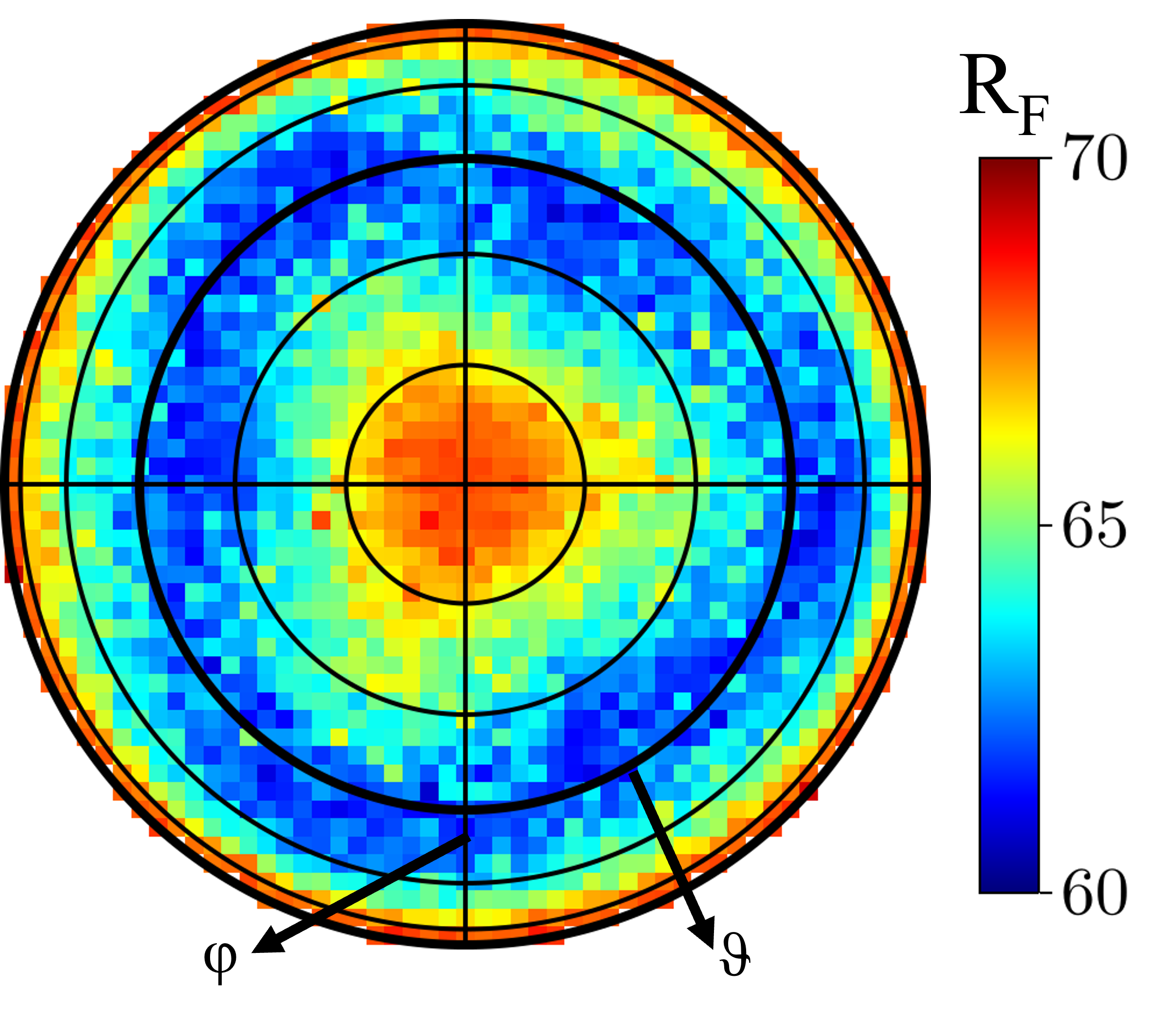}
        \caption{A contour plot representing the relationship between the \ch{R_{F}} factor of fitting and the direction of the rotation axis in the spiral model. The spherical coordinates are mapped to polar coordinates by using Lambert azimuthal equal-area projection.}
        \label{fig:ernpdc}
    \end{figure}  
    
    \subsection{\ch{Er_{2}O_{2}CO_{3}}}
    
        For \ch{Er_{2}O_{2}CO_{3}}, the high temperature CW fit of \sfrac{1}{$\chi$} (Fig. \ref{fig:sus}(g)) yields a $\theta_{\text{CW}}$ = -2.94 K and a $\mu_{\text{eff}}$ = 7.85 $\mu_{\text{B}}$. This $\mu_{\text{eff}}$ value is slightly smaller than the free-ion moment of $\mu_{\text{eff}}^{\text{theo}}$ = 9.58 $\mu_{\text{B}}$ expected for \ch{Er^{3+}} ions. The low temperature CW fit yields a $\theta_{\text{CW}}$ = -4.00 K and a $\mu_{\text{eff}}$ = 7.35 $\mu_{\text{B}}$. 

        For \ch{Er_{2}O_{2}CO_{3}}, the $\chi'(T)$ (Fig. \ref{fig:er}(a)) shows a peak around $T_{\text{N}}$ = 1.58 K, suggesting the occurrence of a long range magnetic ordering. With increasing DC magnetic field, this feature shifts to low temperatures. The $\chi'(\text{B})$ data (Fig. \ref{fig:er}(b)) measured at base temperature $T$ = 0.46 K shows a peak at \ch{B_{C}} = 0.76 T and becomes flat around \ch{B_{S}} = 3.0 T (not shown here).  With increasing temperature, the peak position does not change too much but the peak becomes weaker and disappears above 1.55 K. Its magnetic phase diagram was constructed in Fig. \ref{fig:er}(c).

        NPD measurements of the Er sample were also performed to study its magnetic structure. The NPD pattern measured at 20 K confirms its hexagonal lattice structure with space group $P6_{3}/mmc$ (Fig. \ref{fig:struct}(a)). The lattice parameters were determined to be $a$ = $b$ = 3.823 (1) \AA{} and $c$ = 15.044(2) \AA{}, which are consistent with XRD results (Table \ref{table:rr}). The NPD pattern measured at 0.3 K showed 3 main magnetic peaks (Fig. \ref{fig:ernpdr}(b)). Based on them, the propagation vector was resolved to be (\sfrac{1}{3}, \sfrac{1}{3}, 0.41), indicating incommensurate magnetic ordering. While the exact magnetic structure was unsolvable due to the number of parameters in the refinement---resulting from the four independent magnetic ions in each unit cell---attempts to refine the data were made by introducing several extra constraints.
        
        Using the FULLPROF software, a spiral model was used to represent the spins in the refinement. As shown in Fig. \ref{fig:ernpdm}(a), each spin (green arrow) in a unit cell can be described by three angles. $\theta$ and $\phi$ represent the orientation of the rotation axis, and $\psi$ describes the angle between the spin and the rotation axis. A relative phase factor also needs to be included to determine the orientation of the spin. Thereafter, the spins in all the other unit cells can be generated by rotating the spins around rotation axes according to the propagation vector. To refine the experimental data, several constrains were included by using this spiral model. For the four spins in each unit cell, we assumed (i) all of them have the same magnetic moment and  same rotation axis and (ii) the $\psi$ angle is 90 degrees.  It is obvious that, under these constraints, all spins lie in the plane perpendicular to the rotation axis. Accordingly, the refinement shows that the $\theta$ angle was resolved to be around 90 degrees, which means the rotation axis lies in the $ab$-plane. The goodness of fitting appeared unaffected by the value of $\phi$. To confirm this, the whole ($\theta$,$\phi$) space with a 2 degree step size was surveyed and constructed a (R,$\theta$,$\phi$) contour plot constructed (Fig. \ref{fig:ernpdc}). The sphere was mapped to a disk using the Lambert azimuthal equal-area projection. From the plot, it is confirmed that the fit is optimized as long as $\theta$ is near 90 degrees. This a reasonable result due to the isotropy in the triangular lattice ($ab$-plane).

        Based on all these considerations, refinement of the 0.3 K NPD data leads to a spiral spin structure, as shown in Fig. \ref{fig:ernpdr}(b) and (c). Several features are (i) the magnetic moment of each spin is 4.21(2) $\mu_{\text{B}}$; (ii) the rotation axis can be any orientation within the $ab$-plane--it was chosen to be the $b$-axis in Fig.\ref{fig:ernpdm} for convenience. (iii) the $k_{x}$, $k_{y}$ are both \sfrac{1}{3}, suggesting a 120-degree spin structure in the triangular layer ($ab$-plane); (iv) along the $c$-axis, each spin in the unit cell forms a cycloid structure with the rotation angle to be $k_{z}$*360=147.6 degrees (Fig. \ref{fig:ernpdm}(c)); (v) the two spins in each bilayer (highlighted by the dashed rectangle in Fig. \ref{fig:ernpdm}(b) and (c)) have inversion symmetry relations (from (x,y,z) to (-x,-y,-z)), which requests them to have opposite propagation vectors ($k$ and $-k$) in an incommensurate ordering state. Therefore, these two spins rotate in opposite directions while the angle between them varies in different bilayers; meanwhile, (vi) the inter-bilayer spins share the same propagation vector and that are roughly ferromagnetically aligned. This fit, while imperfect, captures the main features of the NPD pattern. The difference of the magnetic peak intensities between the data and our calculation may originate from the extra constraints we included and the possible diffuse scattering as suggested by the broadening of the magnetic peaks.

        This proposed spin structure was also confirmed to be reasonable by representative analysis. In the space group $\text{P}6_{3}/\text{mmc}$ with the propagation vector  ${\bf k}=( .33333,~ .33333,~ .42319)$,the  decomposition of the magnetic representation for  all the Er sites are  $\Gamma_{\text{mag}}=1\Gamma_{1}^{1}+1\Gamma_{2}^{1}+2\Gamma_{3}^{2}$. And the basis vectors of $\Gamma_{3}$ irreducible representative can best describe this spin structure. 
    \begin{figure*}[htb!]
        \centering
        \includegraphics[width=0.67\columnwidth]{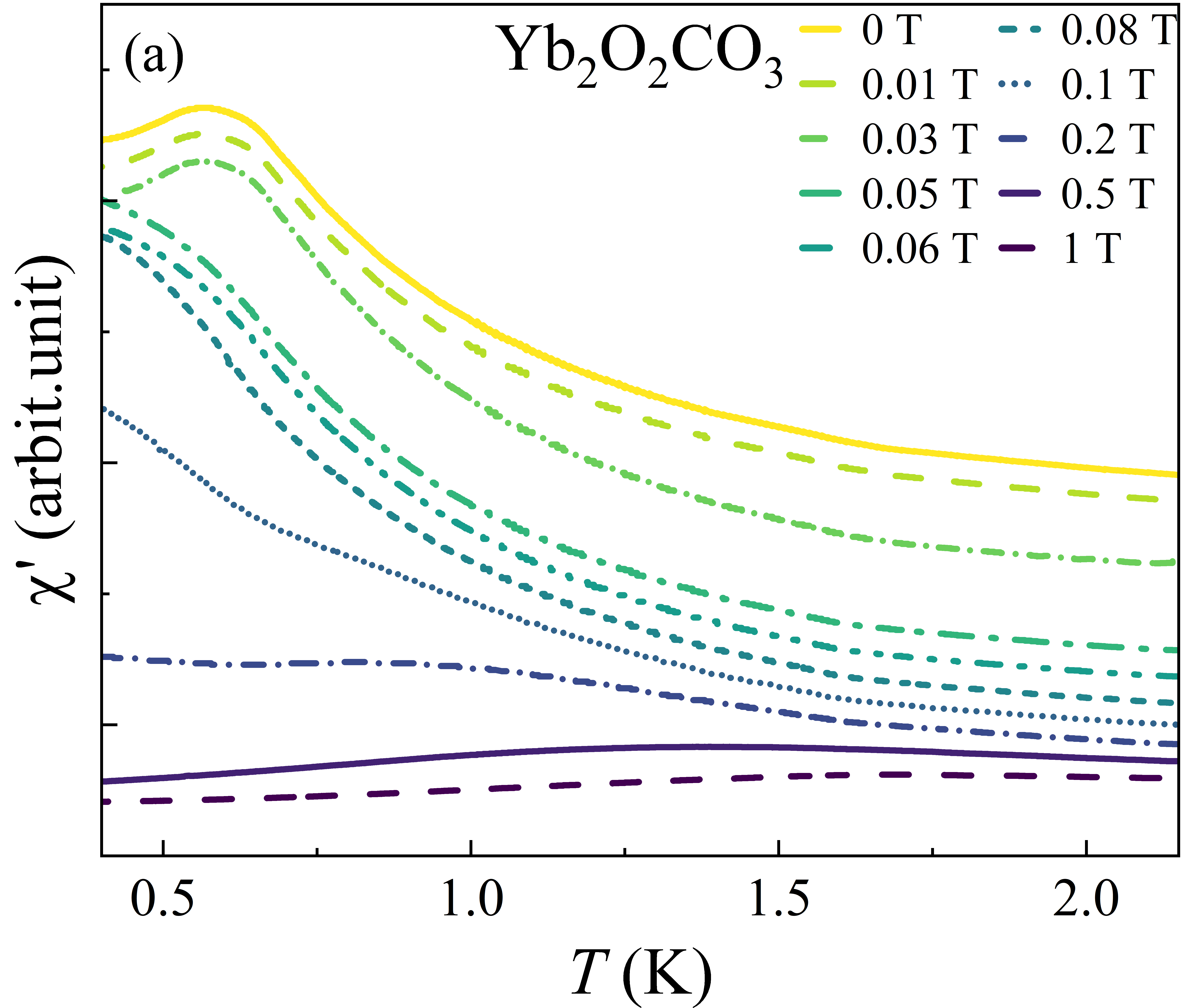}
        \includegraphics[width=0.67\columnwidth]{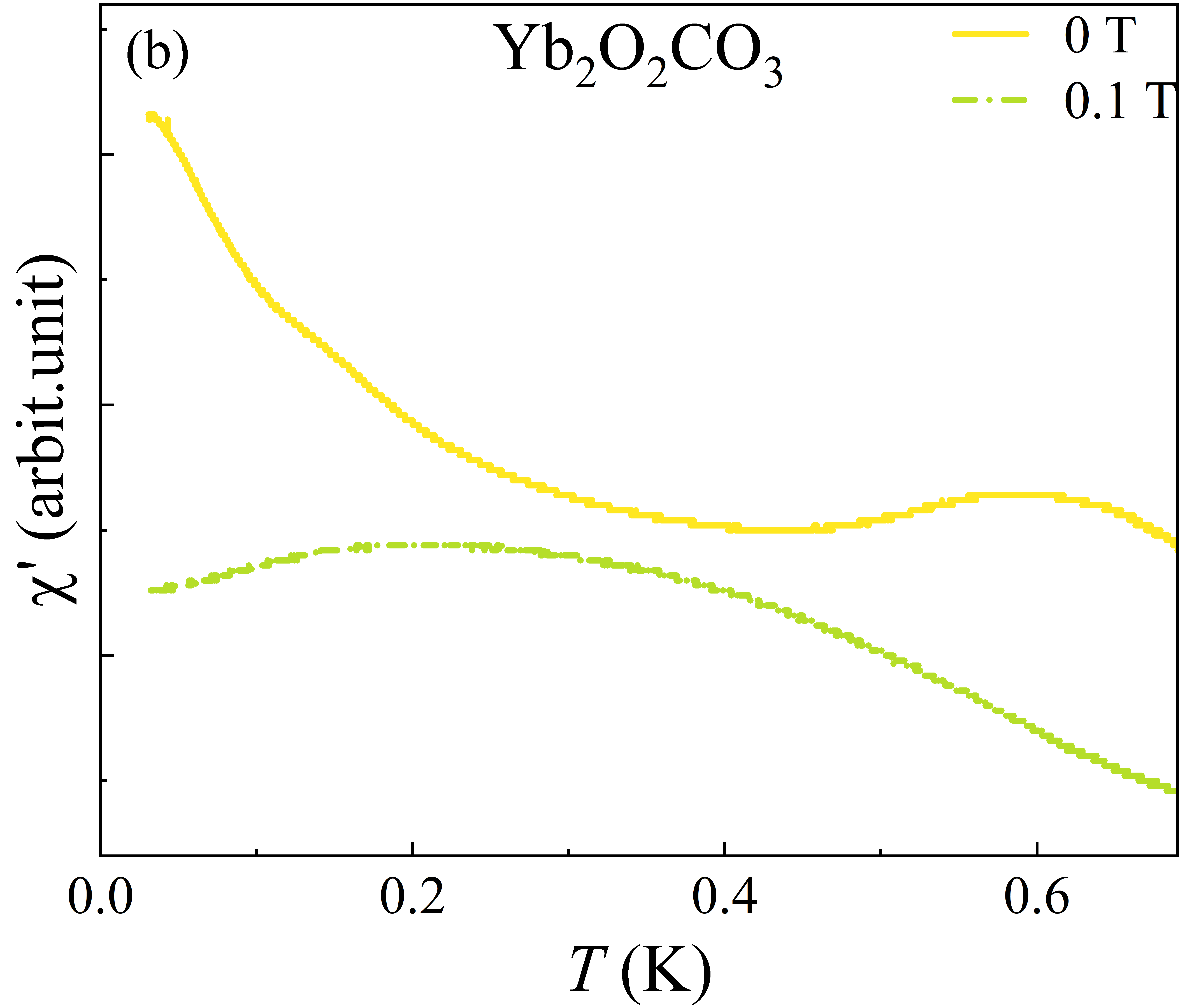}
        \\
        \includegraphics[width=0.67\columnwidth]{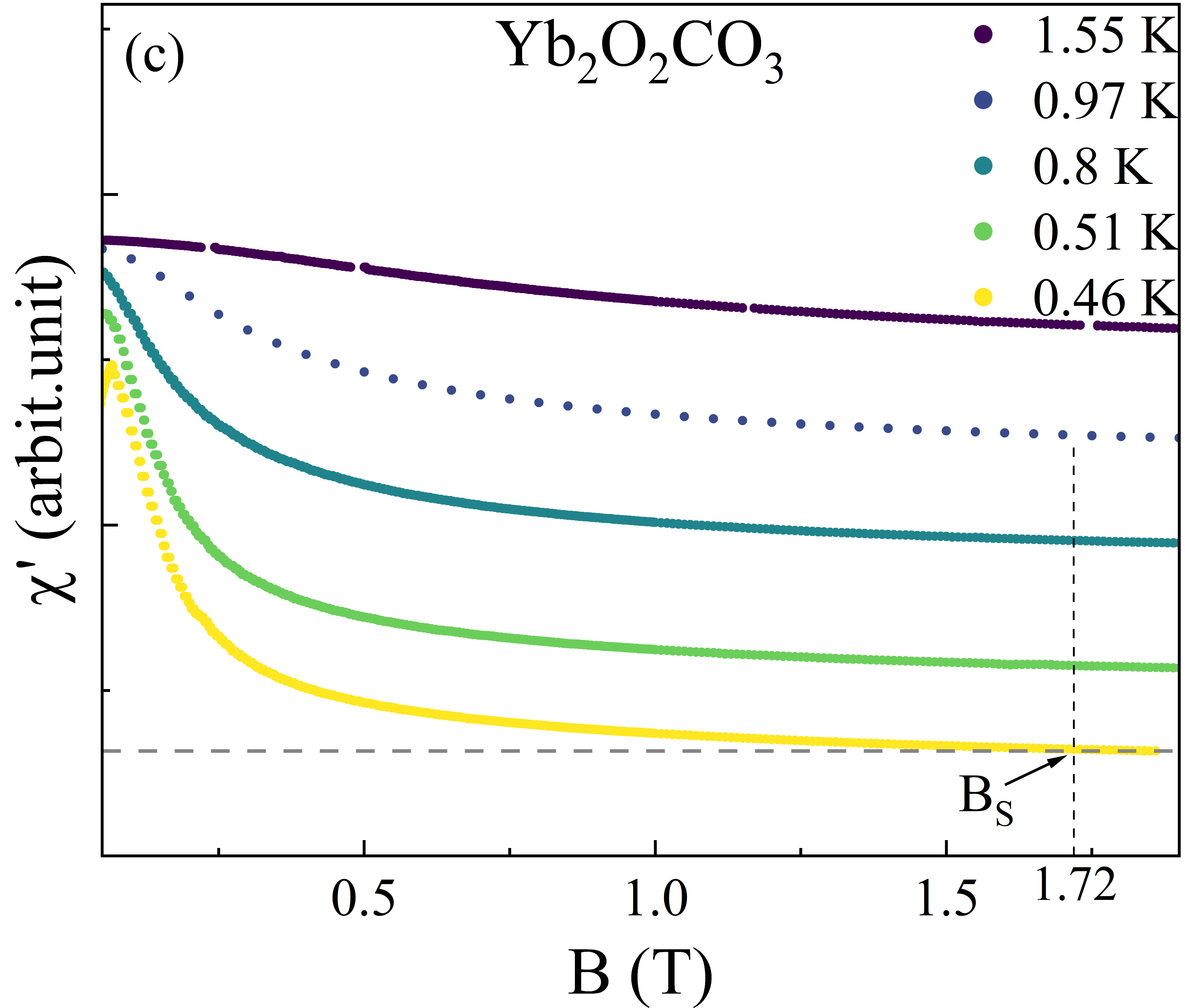}
        \includegraphics[width=0.67\columnwidth]{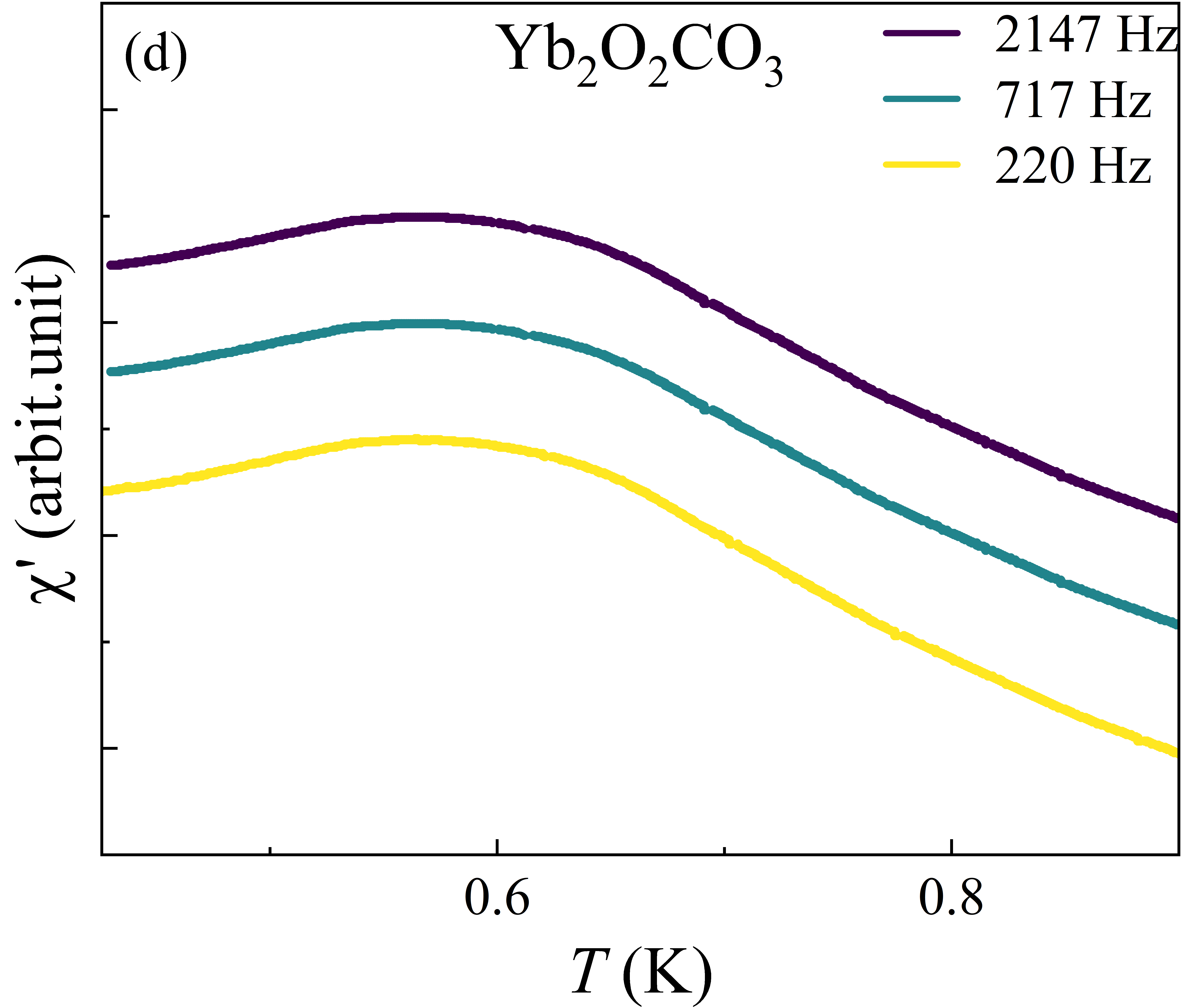}
        \caption{\ch{Yb_{2}O_{2}CO_{3}} (a): Temperature dependence of $\chi'$ measured at various DC fields.. (b): Low temperature regime--down to 50 mK--for zero and 0.1 T data. (c): DC field dependence of $\chi'$ at various temperatures. (d): Temperature dependence of $\chi'$ with different AC field frequencies.}
        \label{fig:yb}
    \end{figure*}
    
    \subsection{\ch{Yb_{2}O_{2}CO_{3}}}
    
        For \ch{Yb_{2}O_{2}CO_{3}}, The high temperature CW fit of \sfrac{1}{$\chi$} (Fig. \ref{fig:sus}(h) yields a $\theta_{\text{CW}}$ = -44.25 K and a $\mu_{\text{eff}}$ = 4.39 $\mu_{\text{B}}$. This $\mu_{\text{eff}}$ value is slightly smaller than the free-ion moment of $\mu_{\text{eff}}^{\text{theo}}$ = 4.54 $\mu_{\text{B}}$ expected for \ch{Yb^{3+}} ions. The low temperature CW fit yields a $\theta_{\text{CW}}$ = -0.94 K and a $\mu_{\text{eff}}$ = 3.25 $\mu_{\text{B}}$. 

        $\chi'(T)$ (Fig. \ref{fig:yb}(a, b)) shows no sharp feature but a broad peak around 0.6 K. With increasing DC magnetic field, this broad peak shifts to low temperatures. For example, it shifts to 0.2 K with B = 0.1 T (Fig. \ref{fig:yb}(b)). The $\chi'(\text{B})$ data (Fig. \ref{fig:yb}(c)) measured at $T$ = 0.46 K becomes flat around \ch{B_{S}} = 1.5 T. As shown in Fig. \ref{fig:yb}(d), this broad peak shows no obvious frequency dependence between 220 Hz and 2147 Hz. Such a broad peak without sharp features suggests that short range magnetic ordering develops around 0.6 K for \ch{Yb_{2}O_{2}CO_{3}} yet no long range ordering appears down to 0.03 K, the lowest temperature measured (Fig. \ref{fig:yb}(b)). The independence of frequency further suggests this short range ordering is not glass-like.
        
\section{\label{sec:level5}Discussion}

\begin{table*}[htb!]
    \caption{\label{table:interaction} Magnetic ordering temperature and calculated exchange interaction ($J_{\text{nn}}$), dipolar interaction (D), and effective exchange interaction ($J_{\text{ex}}$) for \ch{RE_{2}O_{2}CO_{3}} }
    \renewcommand{\arraystretch}{1.5}
    \begin{ruledtabular}
    \begin{tabular}{cccccc}
        Compound & \makecell{$f$ Orbitals \\ (\ch{RE^{3+}})} &\makecell{Ordering \\ Temperature \\ $T_{\text{N}}$  (K)}&\makecell{$J_{\text{ex}}$ (K)}& \makecell{D (K)} & \makecell{$J_{\text{nn}}$ (K)}\\
        \hline
        \ch{Pr_{2}O_{2}CO_{3}} & $4f^{2}$ & --- &---&---&  ---\\
        \ch{Nd_{2}O_{2}CO_{3}} & $4f^{3}$ & 1.37&--2.75&0.07&  --2.82\\
        \ch{Gd_{2}O_{2}CO_{3}} & $4f^{7}$ & 1.26&--3.61&0.84&  --4.45\\
        \ch{Tb_{2}O_{2}CO_{3}} & $4f^{8}$ & 2.10&--4.46&0.83&  --5.29\\
        \ch{Dy_{2}O_{2}CO_{3}} & $4f^{9}$ & 1.70&--5.80&0.94&  --6.74\\
        \ch{Ho_{2}O_{2}CO_{3}} & $4f^{10}$ & 0.73&--4.56&0.97&  --5.53\\
        \ch{Er_{2}O_{2}CO_{3}} & $4f^{11}$ & 1.58 &--2.67&0.60&  --3.27\\
        \ch{Yb_{2}O_{2}CO_{3}} & $4f^{13}$ & ---&--0.63&0.12&  --0.75\\
    \end{tabular}
    \end{ruledtabular}
\end{table*}

   All \ch{RE_{2}O_{2}CO_{3}} members show antiferromagnetic exchange interactions as shown by the negative $\theta_{\text{CW}}$ obtained by low temperature CW fit. Meanwhile, various magnetic ground states were observed, including (i) a nonmagnetic ground state for the Pr sample; (ii) long range magnetic ordering states for the Nd, Gd, Tb, Dy, Ho, and Er samples; and, (iii) a short range ordering state for the Yb sample. Some general magnetic properties among the long range ordered samples are: (i) for the Nd, Gd, Tb, Dy, and Ho samples, the zero field of $\chi'(T)$ exhibits a broad peak followed by a fast drop, which represents the development of short range ordering correlations and thereafter the occurrence of a long range magnetic ordering, respectively. Such behavior is typical for low dimensional antiferromagnets. (ii) For the Nd, Tb, Dy, Ho, and Er samples, the $\chi'(B)$ data exhibits one or two peaks, which suggests field induced spin state transitions.  
   
   We can categorize the eight \ch{RE_{2}O_{2}CO_{3}} samples by two groups. Five of them (RE = Nd, Gd, Dy, Er, Yb) have Kramers ions, whose single-ion ground state doublet is restrictively protected by time-reversal symmetry, and are thus degenerate in energy for a mean field of zero. Therefore, their low temperature magnetism can be  described by an effective moment $J_{\text{eff}}$ = \sfrac{1}{2}. Among them, \ch{Gd^{3+}} is special due to its hall-filled 4$f$ shell (4$f^{7}$, $S$ = 7/2, $L$ = 0) with Heisenberg-like anisotropy and effective moment $S_{\text{eff}}$ = \sfrac{7}{2}. The remaining three (RE = Pr, Tb, Ho) have non-Kramers ions, whose single-ion ground state could also be a doublet but not necessarily degenerate.

   If the degeneracy of the non-Kramers doublet is naturally removed due to the local environment of the \ch{RE^{3+}} ions, which splits the doublet into two nonmagnetic singlet states with a finite energy difference, two scenarios can occur. First, if the spin-spin interactions (exchange/dipolar interactions) is comparable to the energy splitting of the two lowest singlet states, the spin-spin interactions can act as local exchange fields to mix the two nearby singlet states and recover the magnetic moment. Accordingly, the low temperature magnetism of such a system could be treated with an effective pseudospin $S_{\text{eff}}$ = \sfrac{1}{2}. On the other hand, if the two lowest singlet states get too separated in energy, the spin-spin interactions will be insufficient to induce magnetism and thus a nonmagnetic ground state is expected. Accordingly, the Tb and Ho samples with long range magnetic ordering belong to the first scenario and the Pr sample with a nonmagnetic ground state belongs to the second. Recently, several other Pr-compounds \ch{Pr_{3}A_{2}Sb_{3}O_{14}} (A = Mg and Zn) \cite{sanders:2016,dun:2017} with a \ch{Pr^{3+}} kagome lattice and \ch{Pr_{3}BWO_{9}} \cite{nagl:2024} with a distorted \ch{Pr^{3+}} kagome lattice have also been reported to exhibit a nonmagnetic ground state.

   Indeed, the low temperature magnetic properties of the Nd, Tb, Dy, Ho, Er, and Yb samples, such as that the obtained low temperature $\mu_{\text{eff}}$ is smaller than the free-ion moment and that the measured magnetization at 2 K saturates at a value smaller than the expected free-ion saturation value, reflect the nature of effective spin-\sfrac{1}{2} moment. 

   The effective exchange interaction ($J_{\text{ex}}$) between the nearest neighbor \ch{RE^{3+}} ions on a triangular lattice can be approximately calculated as $J_{\text{ex}}$ = 3$k_{\text{B}}\theta_{CW}/zS(S+1)$, where we use $S$ = \sfrac{1}{2} due to the effective spin-\sfrac{1}{2} moment and $z$ = 6 as the number of nearest-neighbor spins. Here, we neglect the two kinds of inter-layer interactions along the $c$-axis. One is the interaction between the adjacent two bilayers, which should be small due to the large distance and the blockage of carbonate group between them. The other is the interaction within each bilayer, which again should be small since the relative shift of the \ch{RE^{3+}} ions between two triangular layers within each bilayer leads to a frustrated inter-layer magnetic coupling. Meanwhile, the dipolar interaction can be estimated by $D$ = $\mu_{0}\mu_{eff}^{2}/4\pi RE_{nn}^{3}$, where \ch{RE_{nn}} is the nearest neighbor distance of \ch{RE^{3+}} ions on a triangular layer. Since the effective exchange interaction $J_{\text{ex}}$ = $J_{\text{nn}}$ + $D$, in which $J_{\text{nn}}$ is the magnetic superexchange interaction between the nearest neighbor \ch{RE^{3+}} ions, we calculated the $J_{\text{nn}}$ value as listed in Table \ref{table:interaction}.

   It is interesting to observe only short range ordering but no long range ordering down to 0.03 K for \ch{Yb_{2}O_{2}CO_{3}}, which makes it another possible QSL candidate with $J_{\text{eff}}$ = \sfrac{1}{2} \ch{Yb^{3+}}-triangular lattice. To compare it with other reported Yb-TLAF QSL candidates, their $\theta_{\text{CW}}$ and $\mu_{\text{eff}}$ obtained from the low temperature CW fit and nearest-neighbor distance of \ch{Yb^{3+}} ions are listed in Table \ref{tab:dist}. While the strength of the exchange interaction represented by the $\theta_{\text{CW}}$ does not simply depend on the nearest neighbor distance but also the exchange path which varies in different systems, one general trend is that the $\theta_{\text{CW}}$ notably decreases after the nearest neighbor distance exceeds 5 \AA{}. Such a small exchange interaction makes them not ideal systems to explore exchange interaction based QSL state. For \ch{Yb_{2}O_{2}CO_{3}}, a $\theta_{\text{CW}}$ = -0.94 K represents a moderate strong exchange interaction compared to the others, which makes it possible to study a QSL state below the Kelvin energy level. As mentioned above, the disrupted inter-layer interaction within the bilayer could be the main obstacle of long range ordering in \ch{Yb_{2}O_{2}CO_{3}}. A similar example is the 6HB-\ch{Ba{_3}NiSb{_2}O{_9}} with \ch{Ni^{2+}}-triangular lattice, in which the adjacent \ch{Ni^{2+}} layers have the same shift as those of \ch{Yb_{2}O_{2}CO_{3}} \cite{cheng:2011,quilliam:2016}. 6HB-\ch{Ba{_3}NiSb{_2}O{_9}} also exhibits no magnetic ordering down to at least 0.3 K and has been studied as a rare spin-1 QSL candidate. Meanwhile, 6HA-\ch{Ba{_3}NiSb{_2}O{_9}}, another phase of \ch{Ba{_3}NiSb{_2}O{_9}} with the \ch{Ni^{2+}}-triangular layers arranged exactly on top of one another, displays long range magnetic ordering at 13.5 K, for which the inter-layer coupling plays an important role \cite{doi:2004,cheng:2011,shirata:2011}. Both \ch{Yb_{2}O_{2}CO_{3}} and 6HB-\ch{Ba{_3}NiSb{_2}O{_9}} demonstrate that TLAFs with shifted triangular layer arrangements can be a good platform for exploring QSL candidates.

   Finally, the magnetic phase diagrams of the Nd and Dy samples show a series of field-induced spin state transitions. When a magnetic field is applied along the easy-axis or within the easy-plane of a TLAF, a spin up-up-down (UUD) phase could be stabilized within a regime of the field to exhibit a \sfrac{1}{3} \ch{M_{s}} magnetization plateau. Such a plateau has been observed in single crystals of TLAFs \ch{Ba_{3}CoSb_{2}O_{9}} \cite{shirata:2012,zhou:2012,susuki:2013,sera:2016,kamiya:2018,mli:2019,fortune:2021}, \ch{Na_{2}BaA(PO_{4})_{2}} (A = Co, Ni, Mn) \cite{kim:2022,li:2021,ding:2021,sheng:2022,li:2020nc}, and \ch{AYbSe_{2}} (A = Na, K, Cs) \cite{ranjith:2019,xing:2021,xing:2019}. Typically, the dM(B)/dB or $\chi'(B)$ data shows two sharp peaks at the critical field positions and a deep valley between them for the UUD phase. For polycrystalline samples, these features become a shallow valley within two broad peaks due to the powder averaging effect, which have been observed in \ch{A_{3}CoB_{2}O_{9}} (A = Ba, Sr and B = Nb, Ta) \cite{lee:2014,lal:2023,ranjith:2017,lee:2017} and \ch{Ba_{2}La_{2}ATe_{2}O_{12}} (A = Co, Ni) \cite{kojima:2018,saito:2019}. A similar phenomena in $\chi'(B)$ was observed for our polycrystalline Nd and Dy samples, which suggests a field induced UUD phase between the two critical fields (Fig. \ref{fig:nd}(d) and Fig. \ref{fig:dy}(c)). Indeed, the magnetization value at \ch{B_{center}} = 1.82 T for Nd and \ch{B_{center}} = 1.42 T for Dy are 0.49 $\mu_{B}/$\ch{Nd^{3+}} and 2.37 $\mu_{B}$/\ch{Dy^{3+}} respectively, which are roughly \sfrac{1}{3} of their \ch{M_{12T}} value, as shown in Fig. \ref{fig:mag}(b),(e). Here \ch{B_{center}} is the field position at the center of the valley of $\chi'(B)$ data, as labeled in Fig. \ref{fig:nd}(c) and Fig. \ref{fig:dy}(b).

\begin{table}
    \begin{threeparttable}
        \caption{\label{tab:dist} Table of Yb-TLAF QSL candidates with their $\theta_{\text{CW}}$ and $\mu_{\text{eff}}$ obtained from the low temperature CW fit and nearest-neighbor distance of \ch{Yb^{3+}} ions. Parentheticals include magnetic field orientations, with ($\perp$) and ($\parallel$) identifying field applied perpendicular and parallel to the \textit{ab}-plane, respectively.}
        \begin{ruledtabular}
        \begin{tabular}{ccccc}
            Compound & \makecell{Low \textit{T} \\ $\theta_{\text{CW}}$ \\ (K)}& \makecell{Low \textit{T} \\ $\mu_{\text{eff}}$ \\ ($\mu_{\text{B}}$)}& \makecell{Nearest- \\ Neighbor \\ Distance \\ (\AA)} &\makecell{Reference}\\
            \hline
            \ch{NaYbO_{2}} & --5.64& 2.84& 3.34&\cite{ding:2019}\\
            \ch{YbZn_{2}GaO_{5}}& --5.22 ($\perp$)& 4.36\tnote{*}& 3.37&\cite{xu:2023}\\
                                & --3.77 ($\parallel$)& ---& &\\
            \ch{YbMgGaO_{4}} & --4.11& 2.8& 3.4 & \cite{li:2015Nat,li:2015PRL}\tnote{\textdagger}\\
            \ch{YbBO_{3}} & --0.8&3.2& 3.75 &\cite{somesh:2023}\\
            \textbf{\ch{Yb_{2}O_{2}CO_{3}}}& \textbf{--0.94}& \textbf{3.25}& \textbf{3.77}&\textbf{This work.}\\
            \ch{NaYbS_{2}} & --13.5 ($\perp$)&3.2 ($\perp$)&  $\sim$ 3.90&\cite{baenitz:2018}\\
                            & --4.5 ($\parallel$)& 1.8 ($\parallel$)& &\\
            \ch{NaYbSe_{2}} & $\sim$ --7&2.43&  4.06&\cite{ranjith:2019}\\
            \ch{CsYbSe_{2}} & --22.6 ($\perp$)& 3.21 ($\perp$)& 4.15&\cite{xing:2020}\\
                            & --13.2 ($\parallel$)& 3.48 ($\parallel$)& &\\
            \ch{NaBaYb(BO_{3})_{2}} & --0.069&2.23&  5.33&\cite{guo:2019}\tnote{\S}\\
            \ch{KBaYb(BO_{3})_{2}} & --0.84&2.67&  5.42&\cite{sanders:2017}\\
            \ch{K_{3}Yb(VO_{4})_{2}} & $\sim$ --1& 2.41& 5.85&\cite{voma:2021}\\
            \ch{Ba_{6}Yb_{2}Ti_{4}O_{17}}  & --0.49& 2.5& 5.91 &\cite{khatua:2024}\\
            \ch{Ba_{3}Yb(BO_{3})_{3}}& --0.077&2.31&  7.18&\cite{cho:2021}\\
        \end{tabular}
        \end{ruledtabular}
        \begin{tablenotes}
            \item[*] Effective moment from high temperature range, 200--300 K.
            \item[\textdagger] Taken from temperatures $\gtrsim8$ K.
            \item[\S] Values averaged from magnetic fields applied perpendicular and parallel to the \textit{c}-axis.
        \end{tablenotes}
    \end{threeparttable}
\end{table}

\section{\label{sec:level6}Conclusion}

    In summary, most of the \ch{RE_{2}O_{2}CO_{3}} (RE = Nd, Gd, Tb, Dy, Ho, Er), either Kramers ions or non Kramers ions, exhibit long range magnetic ordering and field induced spin state transition at low temperatures. Two anomalies are Pr and Yb samples. For Pr sample, its nonmagnetic ground state could be due to the larger gap between the splitting non Kramers ground doublet compared to its spin spin interactions. For Yb sample, the shift between the \ch{Yb^{3+}} ions on adjacent triangular layers within the bilayer causes the frustrated inter-layer interaction and prevents its long range magnetic ordering. This makes \ch{Yb_{2}O_{2}CO_{3}} a good candidate for QSL studies, joining a good number of other \ch{Yb^{3+}}-TLAFs that have been studied as QSL candidates. Certainly, \ch{RE_{2}O_{2}CO_{3}} is a new family of RE-TLAFs exhibiting interesting magnetic properties and deserves future studies.
    
\begin{acknowledgments}
    Research at the University of Tennessee is supported by the Air Force Office of Scientific Research under grant no. FA9550-23-1-0502. A portion of this work was performed at the National High Magnetic Field Laboratory, which is supported by National Science Foundation Cooperative Agreement No. DMR-1644779 and the State of Florida.
\end{acknowledgments}

\bibliography{ref}

\end{document}